\documentclass{aa}  

\usepackage{graphicx}
\usepackage{txfonts}
\usepackage{hyperref}
\usepackage{url}
\usepackage{hhline}
\usepackage{float}
\usepackage{physics}
\usepackage{amsmath}
\usepackage{subcaption}

\newcommand{\msol}{M$_\odot$\,}
\newcommand{\rsol}{R$_\odot$}

\newcommand{\porb}{$P_{\rm orb}$}

\begin{document}

   \title{Observing Orbital Decay in the Ultracompact Hot Subdwarf Binary System ZTFJ213056.71+442046.5}

   \subtitle{}

   \author{Paul Teckenburg\inst{1,5}\fnmsep\thanks{Corresponding author; \texttt{plteckenburg@gmail.com}},
          Thomas Kupfer\inst{1,2},
          Alex J. Brown\inst{1},
          Martin M. Roth\inst{3,4,5},
          Fatma Ben Daya\inst{1},
          Jörg Knoche\inst{1},
          Ananthu K. Lali\inst{6},
          Stella Vje\v{s}nica\inst{5},
          Pa\v{s}ko Roje\inst{5},
          Mike Kretlow\inst{5,7},
          Stefan Cikota\inst{8}}

   \institute{Hamburger Sternwarte, University of Hamburg, Gojenbergsweg 112, 21029 Hamburg, Germany
              \and 
Department of Physics and Astronomy, Texas Tech University, Lubbock, TX 79409, USA
\and
Leibniz Institute for Astrophysics Potsdam (AIP), An der Sternwarte 16, 14482 Potsdam, Germany
\and
Institut für Physik und Astronomie, Universität Potsdam, Karl-Liebknecht-Str. 24/25, 14476 Potsdam, Germany
\and
Deutsches Zentrum für Astrophysik, Postplatz 1, 02826 Görlitz, Germany
\and
Space Science Department, University of Alabama in Huntsville, 320 Sparkman Drive, Huntsville, AL 35899, USA
\and
Instituto de Astrof\'{i}sica de Andaluc\'{i}a (IAA-CSIC), Glorieta de la Astronom\'{i}a, s/n, 18008 Granada, Spain
\and
Centro Astronónomico Hispano en Andalucía, Observatorio de Calar Alto, Sierra de los Filabres, E-04550 Gérgal, Spain
              }

   \date{Received - ; accepted -}

% \abstract{}{}{}{}{} 
% 5 {} token are mandatory
 
  \abstract
   {Ultracompact Galactic binary systems emit low-frequency gravitational waves in the mHz regime. The emission of gravitational waves causes these systems to lose angular momentum, which is detectable by observing the decay of the orbital period of the binary. The system ZTFJ213056.71+442046.5 (ZTF\,J2130) is an ultracompact binary with an orbital period of 39.34 minutes consisting of a Roche lobe-filling hot subdwarf and a white dwarf companion.}
   {We attempt to measure the orbital decay rate $\dot{P}$ caused by gravitational wave emission of ZTF\,J2130 and predict the expected gravitational wave signal for {\it LISA}.}
   {High-speed photometry was conducted using the \texttt{Finger-Lakes-Instrumentation Kepler KL4040FI} Complementary Metal Oxide Semiconductor (CMOS) camera, mounted to the 1.2-meter Oskar Lühning telescope at the Hamburg Observatory as well as the \texttt{Hamamatsu ORCA-Quest\,2} qCMOS camera at the 1.23-meter telescope at the Centro Astronómico Hispano en Andalucía (CAHA) in Spain. ZTF\,J2130 was observed on six nights between August 2024 and September 2025. The obtained lightcurves combined with previous high-cadence observations were used to conduct an $O-C$ timing analysis. Additionally, we employed the {\it LISA} data analysis tool {\sc ldasoft} to model the expected gravitational wave data.}
   {We measure a rate of period change of  $(-2.05\pm0.29)\times10^{-12}\,\mathrm{ss^{-1}}$. Assuming only gravitational wave emission, the rate of period change corresponds to a chirp mass of $(0.42\pm0.04)\,\mathrm{M_{\odot}}$. From {\sc ldasoft} we predict that {\it LISA} will be able to measure the chirp mass with an uncertainty of $\sim10\,\%$. }
   %and a gravitational wave characteristic strain of $(8.13\pm0.20)\times10^{−21}$ after three months of LISA operation. This value and the GW frequency of the object exclude ZTFJ2130 as a verification binary. The proposed lifespan of LISA of four years and the expected extended lifespan of ten years would lead to characteristic strains of $(3.25\pm0.08)\times10^{−20}$ and $(5.14\pm0.12)\times10^{−20}$, respectively, increasing the probability of a detection by LISA.}
   {This work measures the orbital decay with an uncertainty of $\approx14\,\%$ and shows that modern (q)CMOS detectors are well suited to provide precise timing measurements, enabling the measurement of the orbital decay of compact Galactic binaries with high precision even with modest size telescopes. The derived orbital decay is fully consistent with predictions from spectral and lightcurve modeling, although the masses are not known with sufficient precision to measure any deviation of the orbital decay from only gravitational waves. We show that future observations with {\it LISA} can potentially provide a deviation from only gravitational wave effects, e.g. due to accretion, if the effect is sufficiently large.}

   \titlerunning{Orbital Decay of ZTF\,J2130}
   \authorrunning{Teckenburg et al.}
   \maketitle
%
%-------------------------------------------------------------------

\section{Introduction}

Hot subdwarf stars are compact hot stars with spectral type B (sdB) or O (sdO) and luminosities below main sequence stars. Most of them are believed to be helium-core-burning stars with masses around 0.5\,\msol\, and radii around 0.1-0.3\,\rsol \citep{heb86,heb09,heb16}. A large number of hot subdwarf stars are found in close binaries, with orbital periods of \porb$<10$\,days \citep{max01,nap04a,kup15a,sch22}. For such short-period hot subdwarf binaries, common envelope (CE) ejection is the only likely formation channel followed by the loss of angular momentum due to the radiation of gravitational waves (GWs) \citep{han02,han03,nel10a}.

\begin{figure*}[t]
    \centering
    \begin{subfigure}{\columnwidth}
        \includegraphics[width=\columnwidth]{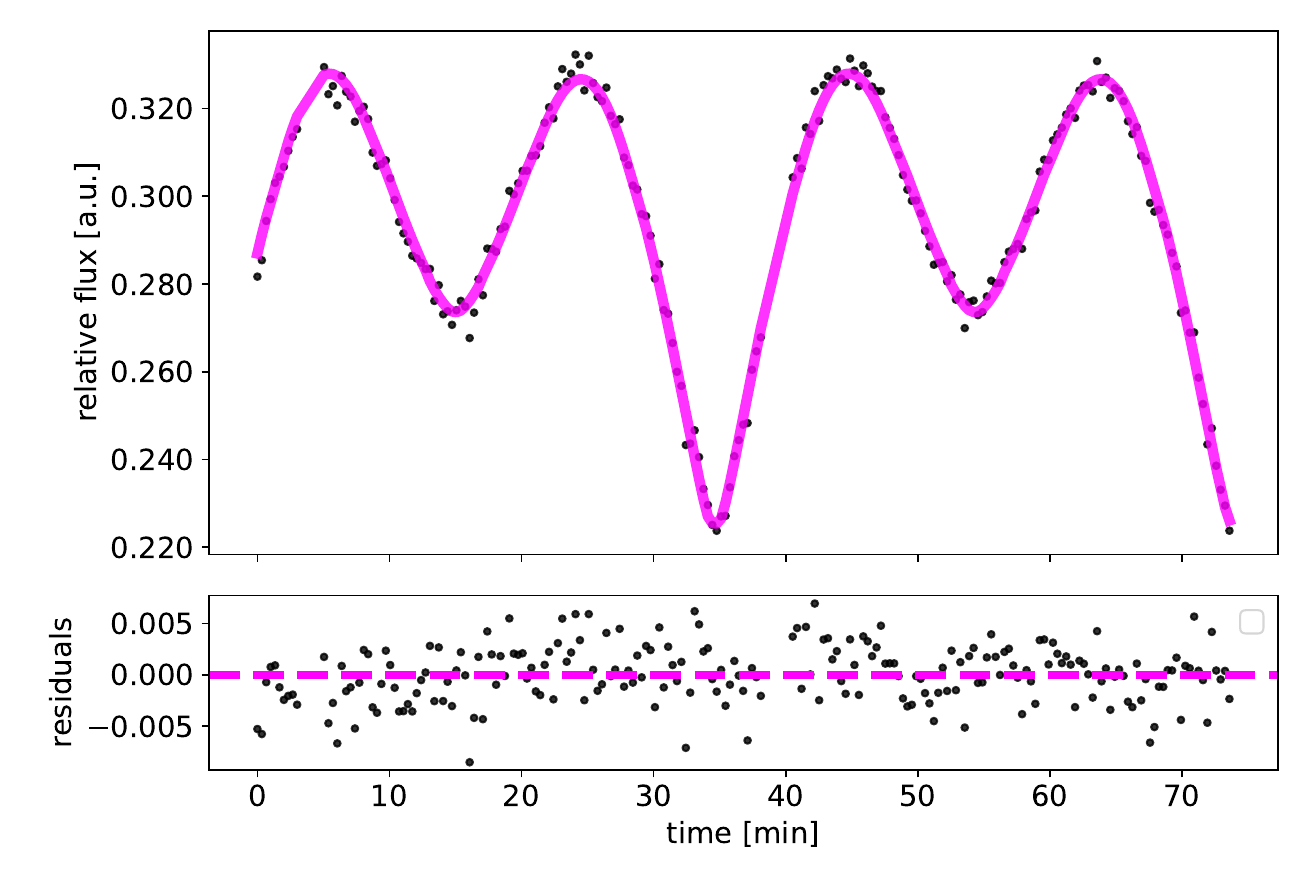}
        \caption{}
        \label{fig:lightcurve_a}
    \end{subfigure}
    \vspace{0.5em}
    \begin{subfigure}{\columnwidth}
        \includegraphics[width=\columnwidth]{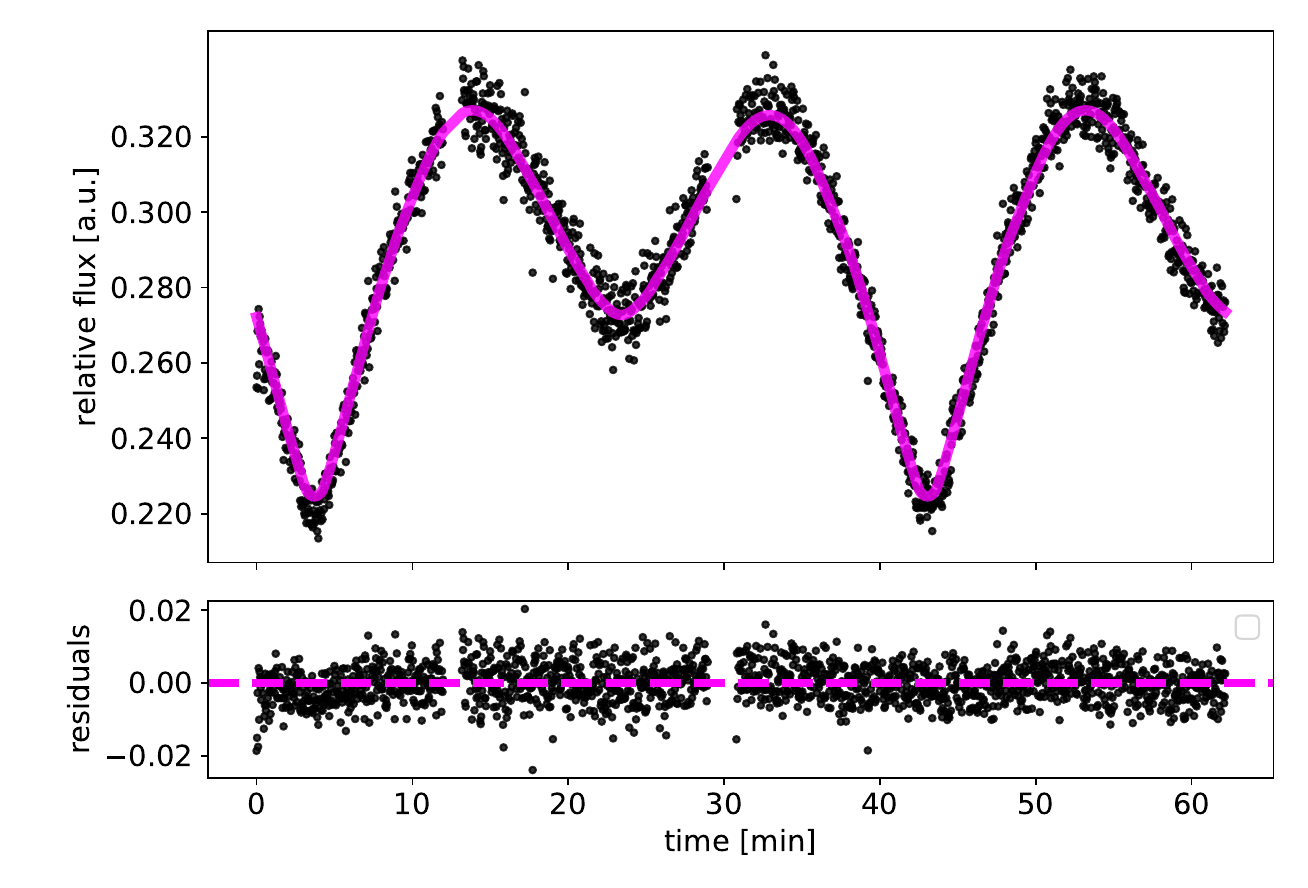}
        \caption{}
        \label{fig:lightcurve_b}
    \end{subfigure}
    \caption{(a) Lightcurve of ZTF\,J2130 taken with the CMOS camera at the OLT in Hamburg on Aug 12, 2024 and (b) lightcurve of ZTF\,J2130 taken with the 1.23-m telescope at CAHA on Sep 24, 2025. Note that $T_{\mathrm{exp,OLT}}\approx 10\times T_{\mathrm{exp,CAHA}}$. Overplotted are lightcurve models generated using \textsc{lcurve}.}
    \label{HC_Lightcurve}
\end{figure*}

%Note that the CAHA exposure times are roughly a factor of ten shorter than the OLT exposure times.

Compact hot subdwarf binaries are found with low-mass main sequence companions and white dwarf (WD) companions (e.g. \citealt{gei11b, gei14, kup15a, sch19, bar22}). Systems with WD companions are of particular interest if the hot subdwarf starts mass transfer of helium rich material before helium is exhausted in the core. After the WD companion accretes $\approx0.1$\,\msol of helium rich material, helium burning is predicted to be ignited unstably in the accreted helium layer on the WD surface \citep{bro15,bau17}. This could either disrupt the WD, even when the mass is significantly below the Chandrasekhar mass, a so-called double detonation supernova  (e.g. \citealt{liv90,liv95,fin10,woo11,wan12,she14,wan18,neu19}) or detonate the He-shell without disrupting the WD, which results in a faint and fast type Ia supernova with subsequent weaker He-flashes \citep{bil07,bro15}.

\citet{bau21} showed that typical hot subdwarf binaries with WD companions which exit the common envelope phase at \porb$\lesssim2$\,hours will reach contact while the sdB is still burning helium. Due to the emission of GWs the orbit of the binary will shrink until the sdB fills its Roche lobe at a period of $\approx30-100$\,min, depending on the evolutionary stage and envelope thickness of the hot subwarf (e.g. \citealt{sav86,tut89,tut90,ibe91,yun08,pie14,bro15,neu19,bau21}).

\begin{table}[!t]
\center
\caption[Observations J2130]{Observations of ZTF\,J2130 with the OLT and CAHA 1.23-m.}
\begin{tabular}{lllll}
\hhline{=====}
Telescope & Date & UT & $N_{\mathrm{exp}}$ & $T_{\mathrm{exp}}$ {[}s{]} \\
\hline
OLT & 2024 Aug 12 & 20:46 - 22:00     & 215  & 20  \\
OLT & 2024 Sep 05 & 19:08 - 21:28     & 552  & 15  \\
OLT & 2024 Nov 30 & 21:43 - 22:29     & 140  & 20  \\
OLT & 2025 Jan 11 & 16:42 - 17:42     & 360  & 10  \\
OLT & 2025 Aug 19 & 20:04 - 20:54     & 200  & 15  \\ 
CAHA & 2025 Sep 24 & 23:45 - 00:47    & 1776 & 2   \\
\hline
\end{tabular}
\label{Observations_J2130}
\end{table}

Upcoming space-based GW detectors, such as the Laser Interferometer Space Antenna {\it LISA} \citep{ama23}, and TianQin \citep{luo16} will be sensitive to GW emission at mHz frequencies from ultracompact binaries. Detailed simulations of Galactic binary populations find that LISA is expected to individually resolve GW signals from $\mathcal{O}(10^4)$ Galactic binaries, while upwards of $\mathcal{O}(10^2)$ of these are also expected to be identified and characterized through their electromagnetic radiation as "multi-messenger" sources \citep{nel04, sha12, sha14a, kor17}. To date, approximately 43 {\it LISA} detectable binaries with orbital periods \porb$<2$\,hours have been characterized through their electromagnetic radiation \cite[see][and references therein]{fin23,kup24}. Currently, only five detached hot subdwarf binaries with a WD companion are known to have \porb$<2$\,hours \citep{ven12,gei13,kup17,kup17a, pel21, kup22}. Just recently \citet{kup20a,kup20} discovered the first two Roche lobe-filling hot subdwarfs, ZTF\,J2055+4651 and ZTF J2130+4420, as part of a high-cadence Galactic Plane survey using the Zwicky Transient Facility \citep{kup21}. ZTF\,J2055+4651 has an orbital period of 56\,min, whereas ZTF J2130+4420 (hereafter ZTF\,J2130) has an orbital period of 39\,min. Due to its short period, ZTF\,J2130 is expected to be a source of GWs, although \citet{kup24} do not list ZTF\,J2130 as one of the {\it LISA} detectable binaries. 

Based on the derived system parameters, \citet{kup20} predict the rate of period change caused by the angular momentum loss due to GW emission to be $\dot{P}=(-1.68\pm0.42)\times10^{-12}\,\mathrm{ss^{-1}}$. Recently, \citet{antipin} found an orbital decay of $\dot{P}=(-2.66\pm0.62)\times10^{-12}\,\mathrm{ss^{-1}}$ in ZTF\,J2130, higher than predicted by \citet{kup20a}, and argued that the increased rate of orbital decay leads to an almost twofold increase in the expected signal-to-noise ratio for the observations of GWs from this binary system with space laser interferometers. Therefore, it is useful to conduct follow-up observations of this system to measure $\dot{P}$ with greater precision.

In this paper, we present follow-up photometric observations of ZTF\,J2130, obtained with the Oscar L\"uhning telescope at the Hamburg observatory and the 1.23-m telescope at the Centro Astronómico Hispano en Andalucía (CAHA), both equipped with new CMOS detectors. We report the detection of a rapid orbital decay rate in the system and discuss the orbital period change in the context of expectations from general relativity. Section \ref{Observations} describes our observations and Section \ref{O-C Timing Analysis} describes in detail the $O-C$ method which was used to measure the orbital decay and the results we obtained. Section \ref{Discussion} discusses the results with respect to predictions from general relativity. We conclude and summarize the paper in Section \ref{Conclusions}.

\begin{figure*}[t]
    \centering
    \includegraphics[width=\hsize]{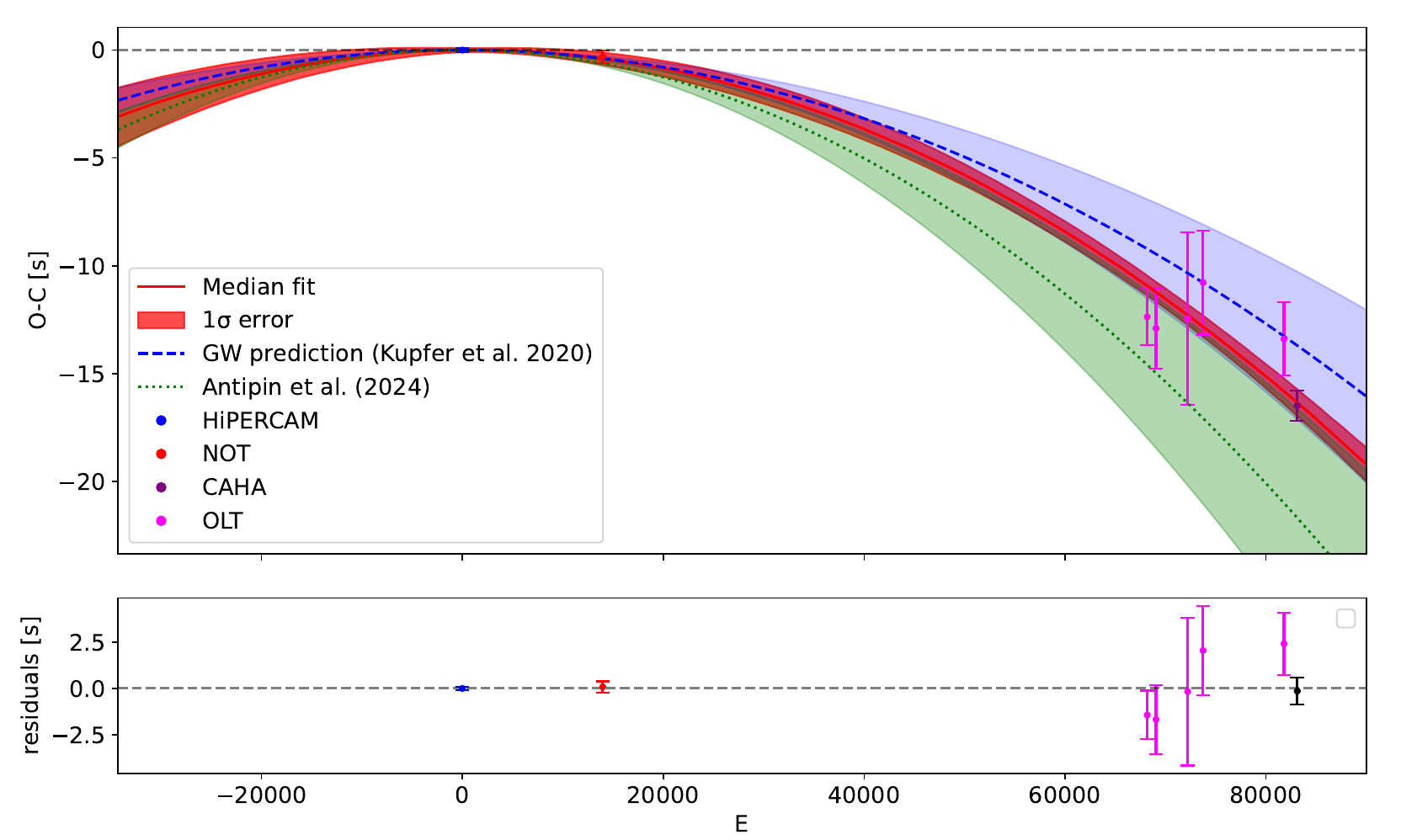}
    \caption{$O-C$ diagram for ZTF\,J2130 with the integer-number of cycles $E$ on the abscissa and the $O-C$ residuals in seconds on the ordinate. Included are data points from the Nordic Optical Telescope (NOT) and the HiPERCAM instrument at the Gran Telescopio Canarias (GTC), taken from \citet{deshmukh}. The new data points have been taken with the OLT in Hamburg and the 1.23-meter telescope at CAHA in Spain. Note that the curves for $E\le0$ cannot be interpreted in a physical way, i.e. as a $\dot{P}\ge0$.}
    \label{O-C}
\end{figure*}
%The data has been reduced with the HiPERCAM pipeline, and a quadratic $O-C$ equation has been fitted to the data.
%--------------------------------------------------------------------
\section{Observations}\label{Observations}

Photometry was obtained with the Oskar Lühning telescope at Hamburg Observatory and with the 1.23-meter telescope at Calar Alto Observatory in Spain. The Oskar Lühning telescope (OLT) is a 1.2-meter Ritchey-Chr\'etien telescope with a focal length of 15.60\,m. It was built in 1975 and was funded by the Lühning foundation. The telescope was mounted on a preexisting mount in a dome from 1954 at the Hamburg Observatory in Hamburg-Bergedorf and remains one of Germany's largest optical telescopes \citep{hunsch}. Since 2024, the telescope has been operated with a new camera which uses a Contemporary Metal Oxide Semiconductor (CMOS) sensor. In front of the camera, an eight-slot filter-wheel is mounted with one slot being empty to allow observations without filter.

The new camera used for the observations with the OLT is a \texttt{Finger-Lakes-Instrumentation Kepler KL4040FI} CMOS camera with a detector size of $4096\times4096$ active pixels and a pixel size of $9\times9\,\mathrm{\mu m}$. With the $8'\times8'$ field of view, this yields a pixel scale of $\sim0.12\,\mathrm{arcsec/pixel}$. The camera uses a scientific CMOS (sCMOS) sensor. This sensor type has multiple advantages over CCDs for high-speed photometry. The camera offers a much faster readout speed than CCDs with up to 24 frames-per-second. Except for very red wavelengths, sCMOS sensors have comparable high quantum efficiency, low readout noise, however higher dark current in comparison to high-quality science-grade CCDs. The \texttt{Kepler KL4040FI} has a quoted readout noise of $3.7e^{-1}$. The signal-to-noise ratio of a sCMOS sensor can outperform CCD detectors, especially a new variant, called qCMOS, with extremely low readout noise for images with short exposure times \citep{roth}. The fast readout and low readout noise are ideal for photometric observations with short exposure times and high timing precision even at modest size telescopes. 

In addition, observations were taken with the \texttt{Hamamatsu ORCA-Quest\,2} qCMOS camera mounted on the 1.23-meter telescope at Calar Alto Observatory. The camera has a detector size of $4096\times2304$ active pixels and a pixel size of $4.6\times4.6\,\mathrm{\mu m}$. The camera has a particularly low readout noise which has been measured to be only $0.3e^{-1}$ and a readout time of only $0.039$\,sec \citep{kry}. We observed using $4\times4$ binning. The focal length of the telescope is about 9870\,mm. resulting in a field of view of $6.6'\times3.7'$ and a pixel scale (unbinned) of $0.10\,\mathrm{arcsec/pixel}$. ZTF\,J2130 was observed without filter on five nights with the OLT and an exposure time of 10-20\,sec as well as on one additional night with the CAHA 1.23-meter telescope and an exposure time of 2\,sec, also without filter. The observations are summarized in Table \ref{Observations_J2130}.

\begin{table}[t]
\centering
\caption[O-C Values J2130 HiPERCAM]{$O-C$ residuals for ZTF\,J2130 from data reduced with the HiPERCAM pipeline for different telescopes, the secondary mid-eclipse time $T_\mathrm{0,obs}$ with its initial error and the scaled error for the reduced $\chi^2$ in seconds and the corresponding number of epoch $E$.}
\begin{tabular}{lllll}
\hhline{=====}
Telescope & $T_\mathrm{0,obs}$\,[TDB] & Err.\,[s] & \textit{O-C}\,[s] & $E$ \\
\hline
HiPERCAM & 2458672.680859 & 0.1 & 0 & 0 \\
NOT      & 2459054.525729 & 0.3 & -0.31 & 13977 \\
OLT      & 2460535.407269 & 1.3 & -12.34 & 68183 \\
OLT      & 2460559.366478 & 1.9 & -12.91 & 69060 \\
OLT      & 2460645.422958 & 4.0 & -12.46 & 72210 \\
OLT      & 2460687.221837 & 2.4 & -10.77 & 73740 \\
OLT      & 2460907.362466 & 1.7 & -13.40 & 81798 \\
CAHA     & 2460943.506149 & 0.7 & -16.48 & 83121 \\
\hline 
\end{tabular}
\label{table:2}
\end{table}

For the OLT data, standard image calibration was conducted following \citet{warner}. Dark and flat frame calibration was conducted using ten dark frames taken on each night with the same exposure times and at the same temperatures as the science frames. Ten twilight flat frames were taken on May 5, 2024 without a filter and with an exposure time of $60\,\mathrm{ms}$ to reach an average number of counts of $75\%$ of the full well depth. To create a master flat frame, 20 dark frames with the same exposure time and the same temperature as the flat frames were also taken on the same night. The resulting master flat frame was used to calibrate all light frames. Image calibration was conducted using a \textsc{Python} pipeline using standard \textsc{astropy} and \textsc{ccdproc} packages \citep{astropy,ccdproc}. A master dark frame was subtracted from the science frames, and the science frames were divided by a master flat frame. Due to the short exposure time and the low noise of the \texttt{Hamamatsu ORCA-Quest\,2} camera, no dark subtraction was performed. There was also no flat frame calibration because no flat frames were available and the weather conditions did not allow us to take twilight flats. 

To extract lightcurves, differential photometry was conducted using the HiPERCAM\footnote{\url{https://github.com/HiPERCAM/}} pipeline, which was slightly modified to be able to extract data from both detectors used in this work \citep{dhi}. The target aperture sizes were scaled with the FWHM fit of the target. The GPS time stamps from both telescopes were converted to barycentric dynamical time (TDB). Photometry extracted from both telescopes is shown in Figure \ref{HC_Lightcurve}.

\section{$O-C$ Timing Analysis}\label{O-C Timing Analysis}

To measure the period change of ZTF\,J2130 due to GW emission, we took the approach of an observed-minus-calculated ($O-C$) analysis. This analysis uses the accumulation of the period change over a long baseline to enable the detection of a small period change \citep{kepler}.

To determine the secondary mid-eclipse time $T_0$, or the time of inferior conjunction of the sdOB, where the WD is eclipsed by the sdOB, we used the lightcurve modeling code \textsc{lcurve} \citep{cop10}. The code uses grids of points to model the two stars. The shape of the stars in the binary is set by a Roche potential. We assume that the orbit is circular and the rotation periods of the stars are synchronized to the orbital period. The flux that each point on the grid emits is calculated by assuming a blackbody of a certain temperature at the bandpass wavelength, corrected for limb darkening, gravity darkening, Doppler beaming, and the reflection effect. $T_0$ is the only parameter that was fitted to every lightcurve separately. All other parameters for \textsc{lcurve} were taken from \citet{kup20} and \citet{deshmukh}, where the initial model for this object was created. We used a Levenberg-Marquardt algorithm for the $T_0$ fit which allowed us to scale the photometric errors on our data points to reach a reduced $\chi^2\approx1$. 

The derived mid-eclipse times $T_{\mathrm{0,obs}}$ are presented in the second column in Table \ref{table:2}. Additionally, we include mid-eclipse times from \citet{deshmukh} taken with high-speed-photometry excluding observations from large scale surveys (e.g. ZTF or ATLAS) where data has been taken at random times over a longer period, making the time-stamps unreliable.

To determine the period change from the observed-minus-calculated values, we need an expression for $O-C$ including a $\dot{P}$ term. The resulting equation is derived as follows.\\
The observed $T_\mathrm{0,obs}$ is assumed to follow the ephemeris equation 
\begin{align}\label{ephemeris}
T_\mathrm{0,obs}(E)=E_\mathrm{0,true}+P_\mathrm{0,true}E
\end{align} 
with the true reference mid-eclipse time $E_{\mathrm{0,true}}$, the true period $P_\mathrm{0,true}$ at $E_\mathrm{0,true}$ and the epoch (number of cycles) $E$. Note that $E_{\mathrm{0,true}}$ has units of time and $E(t)$ is an integer. Assuming that the period is slowly changing over time, Equation \eqref{ephemeris} can now be expanded using a Taylor-expansion at $E=0$:
\begin{align}
T_\mathrm{0,obs}(E(t))&=\eval{T_\mathrm{0,obs}}_{E=0}+\eval{\dv{T_\mathrm{0,obs}}{E}}_{E=0}E+\eval{\dfrac{1}{2}\dv[2]{T_\mathrm{0,obs}}{E}}_{E=0}E^2.
\end{align}
By applying the chain-rule, the second derivative can now be written as
\begin{align}
\dv[2]{T_0}{E}=\dv{P}{E}
=\dv{t}{E}\dv{P}{t}
=P\dv{P}{t}=P\dot{P}.
\end{align}
This yields
\begin{align}\label{Observed_T0}
T_\mathrm{0,obs}&=E_\mathrm{0,true}+P_\mathrm{0,true}E+\dfrac{1}{2}P_\mathrm{0,true}\dot{P}_\mathrm{true}E^2.
\end{align}
Since we do not know the true values, we have to use the observed values here because we want to calculate an unknown quantity.\\
We use the ephemeris equation for the calculated mid-eclipse time with the values of \citet{deshmukh} of
\begin{align}\label{Calculated_T0}
    T_{\mathrm{0,calc}}&=E_{\mathrm{0,obs}}+P_{\mathrm{0,obs}}E\\
        &=2458672.68085911(8)+0.0273195159(7)\times E. \nonumber
\end{align}
Using Equations \eqref{Observed_T0} and \eqref{Calculated_T0}, the observed minus calculated residual can be written as
\begin{align}\label{O-C_1}
O-C&=\underbrace{E_\mathrm{0,true}+P_\mathrm{0,true}E+\frac{1}{2}P_\mathrm{0,true}\dot{P}_\mathrm{true}E^2}_\mathrm{T_\mathrm{0,obs}}-\underbrace{\left(E_\mathrm{0,obs}+P_\mathrm{0,obs}E\right)}_\mathrm{T_\mathrm{0,calc}}\\
&=E_\mathrm{0,true}-E_\mathrm{0,obs}+\left(P_\mathrm{0,true}-P_\mathrm{0,obs}\right)E+\frac{1}{2}P_\mathrm{0,true}\dot{P}_\mathrm{true}E^2.
\end{align}
The true values are unknown to us, so we can use the errors $\delta E_0$ and $\delta P_0$ as the difference between the true values and the observed values. So we can write Equation \eqref{O-C_1} as
\begin{align}\label{O-C_2}
O-C=\delta E_0+\delta P_0E+\frac{1}{2}(P_0+\delta P_0)\dot{P}_\mathrm{true}E^2.
\end{align}
According to this equation, an $O-C\neq0$ could indicate the following things:
\begin{enumerate}
\item A constant vertical offset indicates a difference between the true and the observed mid-eclipse time at $E=0$.
\item A linear trend with constant slope indicates a difference between the true and the observed period.
\item A quadratic trend might indicate a constant rate of change in period $\dot{P}\neq0$. 
\end{enumerate}
The goal of this analysis is to find a value for the rate of period change. This can be achieved by fitting Equation \eqref{O-C_2} to the data, so for every measured $T_\mathrm{0,obs}$, the $O-C$ residual must be determined. \\
To get an $O-C$ value from the data, the calculated value with the assumption of no period change has to be calculated, using the ephemeris equation \eqref{ephemeris}. Therefore, the number of cycles since $E_\mathrm{0,true}$ is needed. Assuming $\delta E_0\ll P_\mathrm{true}$, this number can be determined by dividing the time difference between the observed reference mid-eclipse time and the observed mid-eclipse time by the observed period:
\begin{align}\label{epochs}
E=\dfrac{T_\mathrm{0,obs}-E_\mathrm{0,obs}}{P_\mathrm{obs}}.
\end{align}
The result of Equation \eqref{epochs} is rounded to the nearest integer value. Equation \eqref{Calculated_T0} is now used to calculate the observed (O) minus calculated (C) time difference:
\begin{align}\label{O-C_3}
O-C&=T_\mathrm{0,obs}-E_\mathrm{0,obs}-P_\mathrm{0,obs}E.
\end{align}
The resulting $O-C$ measurements are shown in Table\,\ref{table:2}. 

We now determine the period change by fitting Equation \eqref{O-C_2} to the values in Table \ref{table:2}. All parameters in the equation are fitted, except for the period, which is taken from \citet{deshmukh}. The Python module \textsc{emcee} is used for an MCMC fit with 150 walkers and 3000 steps \citep{for13}. The walkers were initialized randomly in the following regions: 
\begin{align*}
    -1<\delta E_0<1,\\
    -10^{-3}<\delta P<10^{-3},\\
    1\times10^{-11}<\dot{P}<3\times10^{-11}.
\end{align*}

\begin{figure*}[t]
    \centering
    \includegraphics[width=0.75\linewidth]{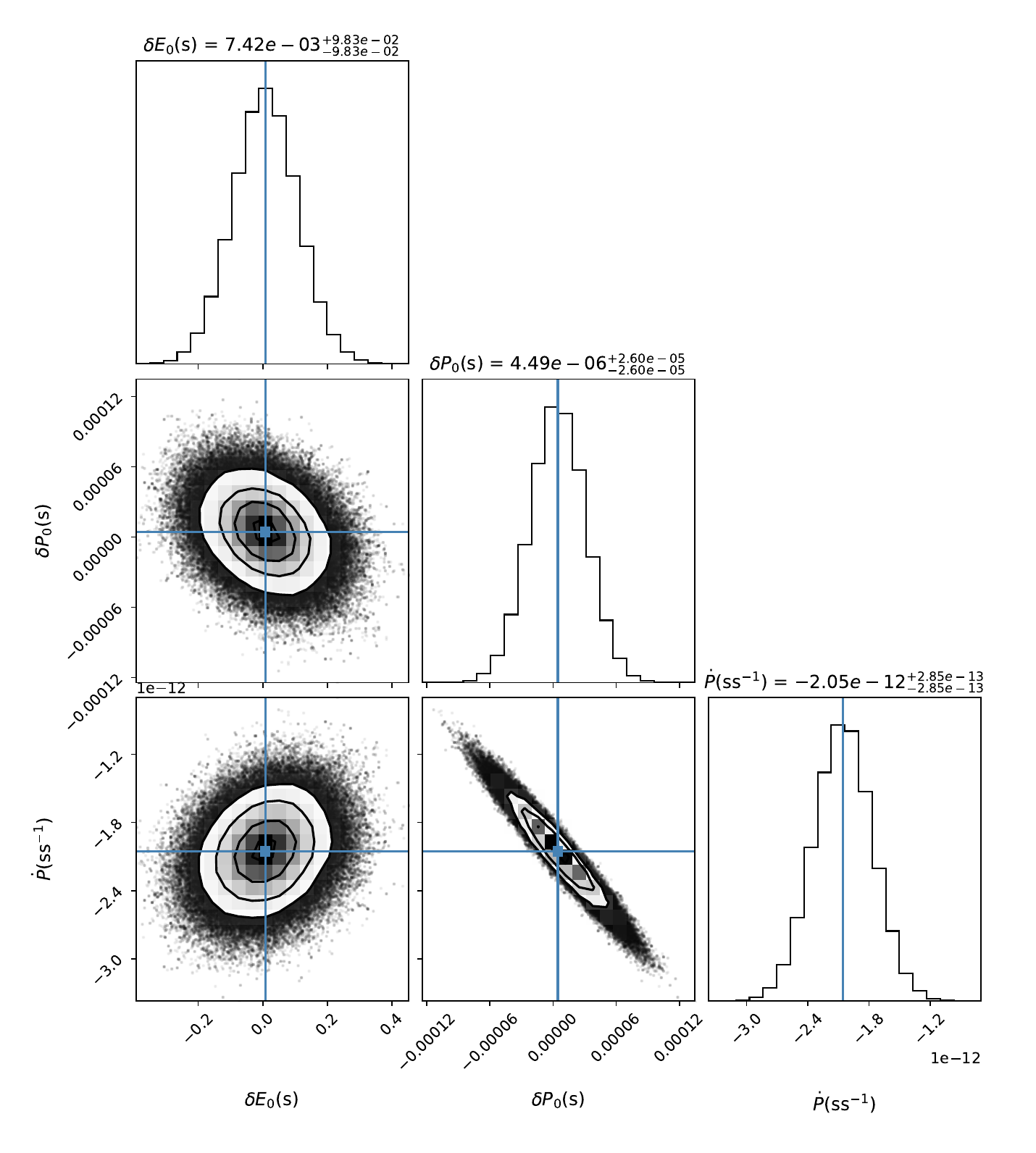}
    \caption{Posterior distribution for the MCMC fit to the $O-C$ data for ZTF\,J2130. The fitted parameters are the error in reference mid-eclipse time $\delta E_0$, the period error $\delta P_0$ and the period decay rate $\dot{P}$.}
    \label{fig:corner}
\end{figure*}

Figure \ref{fig:corner} shows the corner plot for the MCMC fit. The result of the fit is
\begin{align*}
    \delta E_0&=(0.008\pm0.100)\,\mathrm{s}\\
    \delta P_0&=(0.4\pm2.6)\times10^{-5}\,\mathrm{s} \\
    \dot{P}&=(-2.05\pm0.29)\times10^{-12}\,\mathrm{ss^{-1}}.
\end{align*}
The fit is shown in Figure \ref{O-C} along with the prediction by \citet{kup20a} and the measured value from \citet{antipin}.

To determine whether the fit of the model to the data is consistent, we used the \textsc{Python} package \textsc{ConsistencyTEST} by \citet{stoppa}. The consistency test indicates that the $O-C$ data points, the $T_0$ measurement errors, and the fitted $O-C$ model are consistent at a significance level of $\alpha=5\,\%$.

\section{Discussion}\label{Discussion}

%\subsection{\textbf{Consistency of the data with the model}}

\subsection{Comparison to prediction and previous measurements}
The fit results for the errors of the reference mid-eclipse time and the period are $\delta E_0=(0.008\pm0.100)\,\mathrm{s}$ and $\delta P_0=(0.4\pm2.6)\times10^{-5}\,\mathrm{s}$. The values are fully consistent with the values from \citet{deshmukh} of $\delta E_0=\pm0.007\,\mathrm{s}$ and $\delta P_0=\pm6.0\times10^{-5}\,\mathrm{s}$, which are being used in the ephemeris equation \eqref{Calculated_T0}.

\citet{antipin} combined data from the Zwicky Transient Facility with data taken during several nights between April 2023 and August 2024 with the RC600 telescope at the Caucasus Mountain Observatory. Using 17 minima over a total of 6.5 years, they find an orbital decay of $\dot{P}=(-2.66\pm0.62)\times10^{-12}\,\mathrm{ss^{-1}}$. This value is broadly consistent, within the error limits, with predictions from \citet{kup20a} ($\dot{P}=(-1.68\pm0.42)\times10^{-12}\,\mathrm{ss^{-1}}$) which are based on the system parameters derived from spectroscopic and lightcurve modeling assuming angular momentum loss only from GW radiation. Although, \citet{antipin} argue that their measured $\dot{P}$ leads to an almost twofold increase in the expected signal-to-noise ratio for the observations of GWs with {\it LISA}. 

We measure $\dot{P}=(-2.05\pm0.29)\times10^{-12}\,\mathrm{ss^{-1}}$ which is fully consistent with the predictions from \citet{kup20a}. Despite the similar baseline of timing measurements used in this work ($\approx6.5$ years), our uncertainty on $\dot{P}$ is significantly smaller than the value measured by \citet{antipin}. This is due to both the higher SNR and the higher time resolution of our photometry as compared to the ZTF photometry, which dominated their fit. ZTF observed the source irregularly with a low cadence over a longer time frame and used a 30\,sec exposure time. Their individual uncertainty for each $O-C$ measurement is $\approx4-5$\,sec. Our analysis only includes lightcurves that have been taken continuously with a high cadence of 2\,sec to 20\,sec, leading to an uncertainty of $\leq2$\,sec for most $O-C$ measurements, significantly smaller compared to \citet{antipin}. This shows the importance of high cadence observations with little to no readout loss for precise timing measurements. Our work shows that modern (q)CMOS detectors even at modest size telescopes can lead to excellent timing precision needed for precise $\dot{P}$ measurements. Overall, we measure a $\dot{P}$ with an uncertainty of only $14\,\%$, which is fully consistent with the predictions from \citet{kup20a}. Future observations are expected to decrease the uncertainty on $\dot{P}$ even more. With simulated $O-C$ values, following Equation \eqref{O-C_2} and scattered with the typical uncertainty for our measurements ($\approx1$\,sec), we could reach a $\dot{P}$ uncertainty of $\approx1\,\%$ after 10 more years of observation, assuming $\sim4$ observations per year.

\begin{figure}[t]
    \centering
    \includegraphics[width=\hsize]{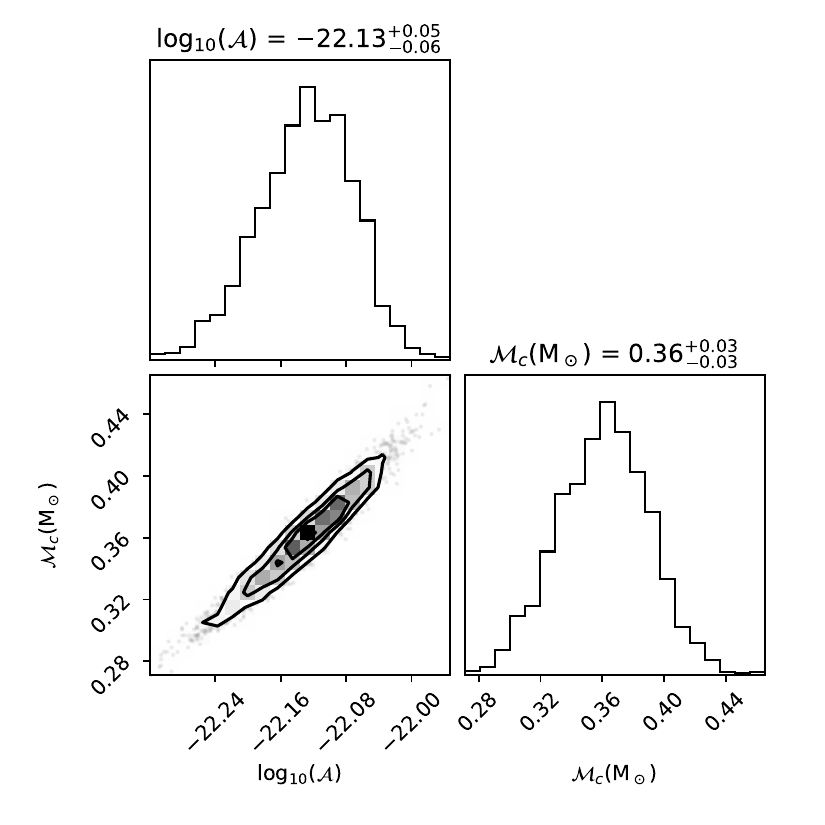}
    \caption{Posterior distribution for the GW amplitude ($\mathcal{A}$) and chirp mass ($\mathcal{M}_c$) for ZTF\,J2130 from gravitational wave observations.}
    \label{fig:lisa}
\end{figure}

%\citet{kupfer} used mass and period measurements from their lightcurves to predict a period decay value of \mbox{$\dot{P}=(-1.68\pm0.42)\times10^{-12}\,\mathrm{ss^{-1}}$.} This value is in full agreement with our value of $\dot{P}=(-1.972\pm0.047)\times10^{-12}\,\mathrm{ss^{-1}}$. 
%\\
%Only one other value has been directly measured for this object. The observations by \citet{antipin} yield a result for the period decay of $\dot{P}=(-2.66\pm0.62)\times10^{-12}\,\mathrm{ss^{-1}}$.

\subsection{Predictions for future space based GW detectors}

At present, projects of space laser interferometers aimed at observing mHz GWs, for example, \texttt{TianQin} \citet{luo16} and the Laser Interferometer Space Antenna ({\it LISA}; \citet{ama23}), are being developed. ZTF\,J2130 has an orbital period of 39\,min and is expected to radiate GWs in the mHz frequency regime. Galactic binaries with well-measured properties are ideal sources for combined electromagnetic and GW multi-messenger studies. In particular, multi-messenger observations can distinguish between the orbital decay coming from GWs and tidal effects or accretion. Assuming an orbital decay only due to angular momentum loss from GWs, we can calculate the chirp mass from 
\begin{equation}\label{equ:chirp_pdot}
\mathcal{M}_c = \frac{c^{3}}{G} \left( \frac{5}{96} \pi^{-8/3} f^{-11/3} \dot{f} \right)^{3/5}
\end{equation}
with $f$ being the GW frequency and $\dot{f}$ being the change in frequency due to angular momentum loss from GW radiation. Using our measured orbital decay, combined with the GW frequency of $f=0.8474$\,mHz we find a chirp mass of $\mathcal{M}_c=0.42\pm0.04\,\mathrm{M_{\odot}}$. Note that this chirp mass assumes angular momentum loss only from GW radiation. Therefore, in the case that other effects such as tides or accretion significantly influence the orbital decay, this chirp mass would not correspond to the actual chirp mass of the binary. From the mass measurements from \citet{kup20a} we can calculate the actual chirp mass following
\begin{equation}
\mathcal{M}_c = \frac{(m_1 m_2)^{3/5}}{(m_1 + m_2)^{1/5}}
\end{equation}
where $m_1=(0.545\pm0.020)\,\mathrm{M_{\odot}}$ is the mass of the accreting white dwarf and $m_2=(0.337\pm0.015)\,\mathrm{M_{\odot}}$ the mass of the hot subdwarf donor. We find a value of $\mathcal{M}_c=0.37\pm0.04\,\mathrm{M_{\odot}}$. This is fully consistent with the chirp mass from the $\dot{P}$ we measured. However, we do not know the masses of the binary well enough from optical observations to obtain a higher precision for $\mathcal{M}_c$ and understand if the orbital decay of ZTF\,J2130 differs from only GW radiation. 

%Tidal effects and accretion predict up to 10\% deviation from GW radiation \citep{ful11,ful12,ful14}.

If GW measurements are in hand, we can derive the actual chirp mass independently following 
\begin{equation}\label{equ:chirp_lisa}
    \mathcal{M}_c = \left[ \frac{\mathcal{A} \, c^4 d}{4 (\pi f)^{2/3} G^{5/3}} \right]^{3/5}
\end{equation}
where $\mathcal{A}$ is the strain amplitude from GW measurements, $f$ is the GW frequency, and $d$ is the distance. The latter two are measured from electromagnetic measurements. As Equation\,\ref{equ:chirp_pdot} assumes orbital decay only from GWs, any deviation of the chirp mass derived with Equation\,\ref{equ:chirp_pdot} from the chirp mass derived with Equation\,\ref{equ:chirp_lisa} points to additional factors in the loss of angular momentum, e.g. due to tidal forces or accretion. 

Both mass transfer and tidal interactions are expected to contribute to the orbital evolution of ZTF~J2130. Assuming conservative mass transfer from the less massive donor to the more massive accretor, the orbital separation evolves as

\begin{equation}
\left(\frac{\dot{a}}{a}\right)_{\mathrm{MT}} = -2(1-q)\frac{\dot{m}_{\mathrm{2}}}{m_{\mathrm{2}}},
\end{equation}

where $q = m_2/m_1$ is the mass ratio \citep{ver88,mar04}. Using the component masses and the estimated mass-transfer rate of $\dot{M}_{\mathrm{2}} \approx 10^{-9}\,M_{\odot}\,\mathrm{yr^{-1}}$ from \citet{kup20a}, we can solve for $\dot{P}_{\mathrm{MT}}$ which corresponds to an expected contribution of 

\begin{equation}
\dot{P}_{\mathrm{MT}} = -3P(1-q)\frac{\dot m_2}{m_2} \approx + 2.5 \times 10^{-13}\,\mathrm{s\,s^{-1}}.
\end{equation}

This corresponds to about $12\%$ of the observed orbital decay rate and is opposite in sign to the effect of gravitational wave emission.

In addition to GW losses and mass transfer, tidal interactions can also contribute to the orbital evolution of ZTF~J2130 by transferring angular momentum between the orbit and stellar spin. For compact hot subdwarf binaries, tidal synchronization timescales are expected to be short compared to the gravitational-wave inspiral timescale \citep{ma24}. Therefore, the donor is likely synchronized with the orbit, consistent with results from \citet{kup20a}. In this regime, tidal torques act primarily to maintain synchronization as the orbital frequency increases, extracting orbital angular momentum at a rate proportional to the stellar moment of inertia. Approximating the donor moment of inertia as $I = km_{\mathrm{2}}R_{\mathrm{2}}^2$ with $k \simeq 0.05$ appropriate for a radiative hot subdwarf (e.g.\ \citealt{sha83, han04}) yields a tidal correction of order

\begin{equation}
\dot{P} \approx \dot{P}_{\mathrm{GW}}\left(1 + \frac{3I_{\mathrm{2}}}{\mu a^{2}}\right) \approx (1\text{--}3)\times10^{-14}\,\mathrm{s\,s^{-1}},
\end{equation}

where $\mu$ is the reduced mass and $a$ the orbital separation \citep{ful11,ful12,ful14}). In contrast, the white-dwarf accretor is expected to be spun up through ongoing accretion and may rotate super-synchronously, in which case its tidal torque does not act as an additional sink of orbital angular momentum and may partially compensate the donor contribution (e.g.\ \citealt{mar04,bil06}). The net tidal contribution is therefore expected to be dominated by the donor and likely modifies the gravitational-wave driven orbital decay only at the percent level. Combined with the expected change in orbital decay from mass transfer, the total expected modification from pure gravitational waves is estimated to be 5-10\% dominated by the mass transfer contribution.

Additionally, the GW amplitude itself might change because of tidal forces or accretion compared to the point mass case. However, \citet{broek} showed this effect to be negligible.

To predict the expected GW signal and the corresponding chirp mass from {\it LISA} and test if we will be able to distinguish an orbital decay only from GWs or additional factors, we employed the parallel-tempered Markov Chain Monte Carlo algorithm, \textsc{ucb\_mcmc} within the \textsc{ldasoft}\footnote{\url{https://github.com/lisa-analysis-center/glass}} package \citep{lit20,glass}. The sky position and orbital period were fixed, while a Gaussian prior was applied to the distance of $d=1.30\pm0.04$\,kpc, which was determined from the Gaia parallax \citep{kup24}, and a uniform prior was applied to the orbital inclination of $i=86.4\pm1.0$\,deg \citep{kup20a}. 

The simulation uses an estimated astrophysical foreground from the unresolved Galactic binaries as described in \citet{cor17}. We ran the simulation for 48 months, consistent with the nominal operation time for {\it LISA}. Following the definition in Section 4 in \citet{kup24}, which states that a source is detected in {\it LISA} when the posterior distribution shows a closed contour, we predict that the source will be detected after 48 months adding ZTF J2130 to the list of detectable {\it LISA} sources when including prior knowledge from EM observations. Using Equation\,\ref{equ:chirp_lisa} we find that {\it LISA} will be able to measure the chirp mass independently from GWs with an uncertainty of $\sim10\,\%$ (see Figure\ref{fig:lisa}).

%After 48 months of LISA observations, we predict an uncertainty for the inclination of $\Delta\iota\approx5^\circ$ and a precision for the GW amplitude ($\mathcal{A}$) of $\sigma_\mathcal{A}/\mathcal{A}\approx26\%$. The chirp mass is expected to be measured with $\mathcal{M} = 0.403\pm0.013$ which is more precise than the current measurement in this work. 

\section{Conclusions and Summary}\label{Conclusions}

In this work, we present follow-up high-speed photometry observations of the ultracompact mass-transferring sdOB+WD binary ZTF\,J2130. Using high-cadence observations obtained with the 1.2\,m OLT equipped with the \texttt{Finger-Lakes-Instrumentation Kepler KL4040FI} CMOS camera at Hamburg Observatory and the 1.23-meter telescope equipped with the \texttt{Hamamatsu ORCA-Quest\,2} qCMOS detector at CAHA combined with data taken with HiPERCAM and the Nordic Optical Telescope we detected an orbital decay of the binary. Using the $O-C$ method, we find an orbital decay of $\dot{P}=(-2.05\pm0.29)\times10^{-12}\,\mathrm{ss^{-1}}$ which is fully consistent with predictions from spectroscopy and lightcurve modeling assuming angular momentum loss only from GWs. Using an observational baseline of 6.5 years we obtain a  $\dot{P}$ measurement with a precision of $\approx{14}\,\%$. We expect that increasing this baseline by another 10 years would enable a $\dot{P}$ measurement to a precision of $\approx1\,\%$. This shows that modern (q)CMOS detectors with low readout noise and virtually no dead time between exposures are well suited for precise timing measurements of compact Galactic binaries even at modest size telescopes. 

Currently, the masses of the binary are not known with sufficient precision to detect any deviation of the orbital decay from only GWs. We employed \textsc{ldasoft} to predict the GW signal from {\it LISA}. Combined with the distance and inclination from \citet{kup20a} we calculate the expected uncertainty on the chirp mass from future {\it LISA} observations and find an expected uncertainty of only $\sim10\%$ which might still improve as the distance uncertainty will decrease with future Gaia data releases. Therefore, any significant deviation of GW angular momentum loss due to e.g. accretion could potentially be detected with future {\it LISA} observations as long as the effect is sufficiently large. 

\begin{acknowledgements}
      This research was supported by Deutsche Forschungsgemeinschaft  (DFG, German Research Foundation) under Germany’s Excellence Strategy - EXC 2121 "Quantum Universe" – 390833306. Co-funded by the European Union (ERC, CompactBINARIES, 101078773). Views and opinions expressed are however those of the author(s) only and do not necessarily reflect those of the European Union or the European Research Council. Neither the European Union nor the granting authority can be held responsible for them.\\
      For this work the HPC-cluster Hummel-2 at University of Hamburg was used. The cluster was funded by Deutsche Forschungsgemeinschaft (DFG, German Research Foundation) – 498394658.\\
      MMR, SV, PR and MK acknowledge support from BMFTR under grant 03WSP1745.\\
      This work is partly based on observations collected at the Centro Astronómico Hispano en Andalucía (CAHA) at Calar Alto, operated jointly by Junta de Andalucía and Consejo Superior de Investigaciones Científicas (CSIC).
\end{acknowledgements}


\bibliographystyle{aa}
\bibliography{refs,refs_1508-2,refs-3}

@misc{lit20,
	author = "Tyson B. Littenberg, Neil J. Cornish",
	title = "GLASS",
	howpublished = "free software (Apache 2.0)",
	year = "2025"
}

@ARTICLE{sch22,
       author = {{Schaffenroth}, V. and {Pelisoli}, I. and {Barlow}, B.~N. and {Geier}, S. and {Kupfer}, T.},
        title = "{Hot subdwarfs in close binaries observed from space. I. Orbital, atmospheric, and absolute parameters, and the nature of their companions}",
      journal = {\aap},
         year = 2022,
        month = oct,
       volume = {666},
          eid = {A182},
        pages = {A182},
          doi = {10.1051/0004-6361/202244214},
archivePrefix = {arXiv},
       eprint = {2207.02001},
 primaryClass = {astro-ph.SR},
       adsurl = {https://ui.adsabs.harvard.edu/abs/2022A&A...666A.182S}
}

@article{roth,
doi = {10.3847/2515-5172/ad90a8},
url = {https://doi.org/10.3847/2515-5172/ad90a8},
year = {2024},
month = {nov},
publisher = {The American Astronomical Society},
volume = {8},
number = {11},
pages = {282},
author = {Roth, Martin M.},
title = {qCMOS Detectors and the Case of Hypothetical Primordial Black Holes in the Solar System, near Earth Objects, Transients, and Other High-cadence Observations},
journal = {Research Notes of the AAS}
}

@article{hunsch,
  title={The telescopes of Hamburg Observatory--history and present situation},
  author={H{\"u}nsch, Matthias},
  journal={Monuments and Sites},
  volume={18},
  pages={274--283},
  year={2009}
}

@article{antipin,
  title={Evolution of the Orbital Period of the Ultracompact Binary System ZTF J213056. 71+ 442046.5},
  author={Antipin, SV and Berdnikov, LN and Postnov, KA and Zubareva, AM and Ikonnikova, NP and Burlak, MA and Belinski, AA},
  journal={Astronomy Letters},
  volume={50},
  number={10},
  pages={619--624},
  year={2024},
  publisher={Springer}
}

@article{deshmukh,
    author = {Deshmukh, Kunal and Kupfer, Thomas and Hakala, Pasi and Bauer, Evan B and Berdyugin, Andrei and Bildsten, Lars and Marsh, Thomas R and Mereghetti, Sandro and Piirola, Vilppu},
    title = {Limiting the accretion disc light in two mass transferring hot subdwarf binaries},
    journal = {Monthly Notices of the Royal Astronomical Society},
    volume = {519},
    number = {1},
    pages = {148-156},
    year = {2022},
    month = {11},
    issn = {0035-8711},
    doi = {10.1093/mnras/stac3492},
    url = {https://doi.org/10.1093/mnras/stac3492},
    eprint = {https://academic.oup.com/mnras/article-pdf/519/1/148/48180354/stac3492.pdf},
}

@book{warner,
  title={A practical guide to lightcurve photometry and analysis},
  author={Warner, Brian D},
  year={2006},
  publisher={Springer}
}

@ARTICLE{kup22,
       author = {{Kupfer}, Thomas and {Bauer}, Evan B. and {van Roestel}, Jan and {Bellm}, Eric C. and {Bildsten}, Lars and {Fuller}, Jim and {Prince}, Thomas A. and {Heber}, Ulrich and {Geier}, Stephan and {Green}, Matthew J. and {Kulkarni}, Shrinivas R. and {Bloemen}, Steven and {Laher}, Russ R. and {Rusholme}, Ben and {Schneider}, David},
        title = "{Discovery of a Double-detonation Thermonuclear Supernova Progenitor}",
      journal = {\apjl},
         year = 2022,
        month = feb,
       volume = {925},
       number = {2},
          eid = {L12},
        pages = {L12},
          doi = {10.3847/2041-8213/ac48f1},
archivePrefix = {arXiv},
       eprint = {2110.11974},
 primaryClass = {astro-ph.SR},
       adsurl = {https://ui.adsabs.harvard.edu/abs/2022ApJ...925L..12K}
}

@ARTICLE{neu19,
       author = {{Neunteufel}, P. and {Yoon}, S. -C. and {Langer}, N.},
        title = "{Evolution of helium star plus carbon-oxygen white dwarf binary systems and implications for diverse stellar transients and hypervelocity stars}",
      journal = {\aap},
         year = 2019,
        month = jul,
       volume = {627},
          eid = {A14},
        pages = {A14},
          doi = {10.1051/0004-6361/201935322},
archivePrefix = {arXiv},
       eprint = {1904.12421},
 primaryClass = {astro-ph.SR},
       adsurl = {https://ui.adsabs.harvard.edu/abs/2019A&A...627A..14N}
}

@ARTICLE{bau21,
       author = {{Bauer}, Evan B. and {Kupfer}, Thomas},
        title = "{Phases of Mass Transfer from Hot Subdwarfs to White Dwarf Companions and Their Photometric Properti\
es}",
      journal = {\apj},
         year = 2021,
        month = dec,
       volume = {922},
          eid = {245},
        pages = {245},
       number = {2},
          doi = {10.3847/1538-4357/ac25f0},
archivePrefix = {arXiv},
       eprint = {2106.13297},
 primaryClass = {astro-ph.SR},
       adsurl = {https://ui.adsabs.harvard.edu/abs/2021arXiv210613297B}
}

@ARTICLE{sha14a,
   author = {{Shah}, S. and {Nelemans}, G.},
    title = "{Constraining Parameters of White-dwarf Binaries Using Gravitational-wave and Electromagnetic Observations}",
  journal = {\apj},
archivePrefix = "arXiv",
   eprint = {1406.3599},
 primaryClass = "astro-ph.IM",
     year = 2014,
    month = aug,
   volume = 790,
      eid = {161},
    pages = {161},
      doi = {10.1088/0004-637X/790/2/161},
   adsurl = {http://cdsads.u-strasbg.fr/abs/2014ApJ...790..161S}
}

@ARTICLE{sha12,
   author = {{Shah}, S. and {van der Sluys}, M. and {Nelemans}, G.},
    title = "{Using electromagnetic observations to aid gravitational-wave parameter estimation of compact binaries observed with LISA}",
  journal = {\aap},
archivePrefix = "arXiv",
   eprint = {1207.6770},
 primaryClass = "astro-ph.IM",
     year = 2012,
    month = aug,
   volume = 544,
      eid = {A153},
    pages = {A153},
      doi = {10.1051/0004-6361/201219309},
   adsurl = {http://adsabs.harvard.edu/abs/2012A%26A...544A.153S}
}

@ARTICLE{kor17,
   author = {{Korol}, V. and {Rossi}, E.~M. and {Groot}, P.~J. and {Nelemans}, G. and 
	{Toonen}, S. and {Brown}, A.~G.~A.},
    title = "{Prospects for detection of detached double white dwarf binaries with Gaia, LSST and LISA}",
  journal = {\mnras},
archivePrefix = "arXiv",
   eprint = {1703.02555},
 primaryClass = "astro-ph.HE",
     year = 2017,
    month = sep,
   volume = 470,
    pages = {1894-1910},
      doi = {10.1093/mnras/stx1285},
   adsurl = {http://adsabs.harvard.edu/abs/2017MNRAS.470.1894K}
}

@ARTICLE{fin23,
       author = {{Finch}, Eliot and {Bartolucci}, Giorgia and {Chucherko}, Daniel and {Patterson}, Ben G. and {Korol}, Valeriya and {Klein}, Antoine and {Bandopadhyay}, Diganta and {Middleton}, Hannah and {Moore}, Christopher J. and {Vecchio}, Alberto},
        title = "{Identifying LISA verification binaries among the Galactic population of double white dwarfs}",
      journal = {\mnras},
         year = 2023,
        month = jul,
       volume = {522},
       number = {4},
        pages = {5358-5373},
          doi = {10.1093/mnras/stad1288},
archivePrefix = {arXiv},
       eprint = {2210.10812},
 primaryClass = {astro-ph.SR},
       adsurl = {https://ui.adsabs.harvard.edu/abs/2023MNRAS.522.5358F}
}

@ARTICLE{ful11,
       author = {{Fuller}, Jim and {Lai}, Dong},
        title = "{Tidal excitations of oscillation modes in compact white dwarf binaries - I. Linear theory}",
      journal = {\mnras},
         year = 2011,
        month = apr,
       volume = {412},
       number = {2},
        pages = {1331-1340},
          doi = {10.1111/j.1365-2966.2010.18017.x},
archivePrefix = {arXiv},
       eprint = {1009.3316},
 primaryClass = {astro-ph.HE},
       adsurl = {https://ui.adsabs.harvard.edu/abs/2011MNRAS.412.1331F}
}

@ARTICLE{ful12,
       author = {{Fuller}, Jim and {Lai}, Dong},
        title = "{Dynamical tides in compact white dwarf binaries: tidal synchronization and dissipation}",
      journal = {\mnras},
         year = 2012,
        month = mar,
       volume = {421},
       number = {1},
        pages = {426-445},
          doi = {10.1111/j.1365-2966.2011.20320.x},
archivePrefix = {arXiv},
       eprint = {1108.4910},
 primaryClass = {astro-ph.SR},
       adsurl = {https://ui.adsabs.harvard.edu/abs/2012MNRAS.421..426F}
}

@ARTICLE{ful14,
       author = {{Fuller}, Jim and {Lai}, Dong},
        title = "{Dynamical tides in compact white dwarf binaries: influence of rotation}",
      journal = {\mnras},
         year = 2014,
        month = nov,
       volume = {444},
       number = {4},
        pages = {3488-3500},
          doi = {10.1093/mnras/stu1698},
archivePrefix = {arXiv},
       eprint = {1406.2717},
 primaryClass = {astro-ph.SR},
       adsurl = {https://ui.adsabs.harvard.edu/abs/2014MNRAS.444.3488F}
}

@ARTICLE{luo16,
       author = {{Luo}, Jun and {Chen}, Li-Sheng and {Duan}, Hui-Zong and {Gong}, Yun-Gui and {Hu}, Shoucun and {Ji}, Jianghui and {Liu}, Qi and {Mei}, Jianwei and {Milyukov}, Vadim and {Sazhin}, Mikhail and {Shao}, Cheng-Gang and {Toth}, Viktor T. and {Tu}, Hai-Bo and {Wang}, Yamin and {Wang}, Yan and {Yeh}, Hsien-Chi and {Zhan}, Ming-Sheng and {Zhang}, Yonghe and {Zharov}, Vladimir and {Zhou}, Ze-Bing},
        title = "{TianQin: a space-borne gravitational wave detector}",
      journal = {Classical and Quantum Gravity},
         year = 2016,
        month = feb,
       volume = {33},
       number = {3},
          eid = {035010},
        pages = {035010},
          doi = {10.1088/0264-9381/33/3/035010},
archivePrefix = {arXiv},
       eprint = {1512.02076},
 primaryClass = {astro-ph.IM},
       adsurl = {https://ui.adsabs.harvard.edu/abs/2016CQGra..33c5010L}
}

@ARTICLE{ama23,
       author = {{Amaro-Seoane}, Pau and {Andrews}, Jeff and {Arca Sedda}, Manuel and {Askar}, Abbas and {Baghi}, Quentin and {Balasov}, Razvan and {Bartos}, Imre and {Bavera}, Simone S. and {Bellovary}, Jillian and {Berry}, Christopher P.~L. and {Berti}, Emanuele and {Bianchi}, Stefano and {Blecha}, Laura and {Blondin}, St{\'e}phane and {Bogdanovi{\'c}}, Tamara and {Boissier}, Samuel and {Bonetti}, Matteo and {Bonoli}, Silvia and {Bortolas}, Elisa and {Breivik}, Katelyn and {Capelo}, Pedro R. and {Caramete}, Laurentiu and {Cattorini}, Federico and {Charisi}, Maria and {Chaty}, Sylvain and {Chen}, Xian and {Chru{\'s}li{\'n}ska}, Martyna and {Chua}, Alvin J.~K. and {Church}, Ross and {Colpi}, Monica and {D'Orazio}, Daniel and {Danielski}, Camilla and {Davies}, Melvyn B. and {Dayal}, Pratika and {De Rosa}, Alessandra and {Derdzinski}, Andrea and {Destounis}, Kyriakos and {Dotti}, Massimo and {Du{\c{t}}an}, Ioana and {Dvorkin}, Irina and {Fabj}, Gaia and {Foglizzo}, Thierry and {Ford}, Saavik and {Fouvry}, Jean-Baptiste and {Franchini}, Alessia and {Fragos}, Tassos and {Fryer}, Chris and {Gaspari}, Massimo and {Gerosa}, Davide and {Graziani}, Luca and {Groot}, Paul and {Habouzit}, Melanie and {Haggard}, Daryl and {Haiman}, Zoltan and {Han}, Wen-Biao and {Istrate}, Alina and {Johansson}, Peter H. and {Khan}, Fazeel Mahmood and {Kimpson}, Tomas and {Kokkotas}, Kostas and {Kong}, Albert and {Korol}, Valeriya and {Kremer}, Kyle and {Kupfer}, Thomas and {Lamberts}, Astrid and {Larson}, Shane and {Lau}, Mike and {Liu}, Dongliang and {Lloyd-Ronning}, Nicole and {Lodato}, Giuseppe and {Lupi}, Alessandro and {Ma}, Chung-Pei and {Maccarone}, Tomas and {Mandel}, Ilya and {Mangiagli}, Alberto and {Mapelli}, Michela and {Mathis}, St{\'e}phane and {Mayer}, Lucio and {McGee}, Sean and {McKernan}, Berry and {Miller}, M. Coleman and {Mota}, David F. and {Mumpower}, Matthew and {Nasim}, Syeda S. and {Nelemans}, Gijs and {Noble}, Scott and {Pacucci}, Fabio and {Panessa}, Francesca and {Paschalidis}, Vasileios and {Pfister}, Hugo and {Porquet}, Delphine and {Quenby}, John and {Ricarte}, Angelo and {R{\"o}pke}, Friedrich K. and {Regan}, John and {Rosswog}, Stephan and {Ruiter}, Ashley and {Ruiz}, Milton and {Runnoe}, Jessie and {Schneider}, Raffaella and {Schnittman}, Jeremy and {Secunda}, Amy and {Sesana}, Alberto and {Seto}, Naoki and {Shao}, Lijing and {Shapiro}, Stuart and {Sopuerta}, Carlos and {Stone}, Nicholas C. and {Suvorov}, Arthur and {Tamanini}, Nicola and {Tamfal}, Tomas and {Tauris}, Thomas and {Temmink}, Karel and {Tomsick}, John and {Toonen}, Silvia and {Torres-Orjuela}, Alejandro and {Toscani}, Martina and {Tsokaros}, Antonios and {Unal}, Caner and {V{\'a}zquez-Aceves}, Ver{\'o}nica and {Valiante}, Rosa and {van Putten}, Maurice and {van Roestel}, Jan and {Vignali}, Christian and {Volonteri}, Marta and {Wu}, Kinwah and {Younsi}, Ziri and {Yu}, Shenghua and {Zane}, Silvia and {Zwick}, Lorenz and {Antonini}, Fabio and {Baibhav}, Vishal and {Barausse}, Enrico and {Bonilla Rivera}, Alexander and {Branchesi}, Marica and {Branduardi-Raymont}, Graziella and {Burdge}, Kevin and {Chakraborty}, Srija and {Cuadra}, Jorge and {Dage}, Kristen and {Davis}, Benjamin and {de Mink}, Selma E. and {Decarli}, Roberto and {Doneva}, Daniela and {Escoffier}, Stephanie and {Gandhi}, Poshak and {Haardt}, Francesco and {Lousto}, Carlos O. and {Nissanke}, Samaya and {Nordhaus}, Jason and {O'Shaughnessy}, Richard and {Portegies Zwart}, Simon and {Pound}, Adam and {Schussler}, Fabian and {Sergijenko}, Olga and {Spallicci}, Alessandro and {Vernieri}, Daniele and {Vigna-G{\'o}mez}, Alejandro},
        title = "{Astrophysics with the Laser Interferometer Space Antenna}",
      journal = {Living Reviews in Relativity},
         year = 2023,
        month = dec,
       volume = {26},
       number = {1},
          eid = {2},
        pages = {2},
          doi = {10.1007/s41114-022-00041-y},
archivePrefix = {arXiv},
       eprint = {2203.06016},
 primaryClass = {gr-qc},
       adsurl = {https://ui.adsabs.harvard.edu/abs/2023LRR....26....2A}
}

@ARTICLE{kepler,
       author = {{Howell}, Steve B. and {Sobeck}, Charlie and {Haas}, Michael and {Still}, Martin and {Barclay}, Thomas and {Mullally}, Fergal and {Troeltzsch}, John and {Aigrain}, Suzanne and {Bryson}, Stephen T. and {Caldwell}, Doug and {Chaplin}, William J. and {et al.}},
        title = "{The K2 Mission: Characterization and Early Results}",
      journal = {PASP},
         year = 2014,
        month = apr,
       volume = {126},
       number = {938},
        pages = {398},
          doi = {10.1086/676406},
archivePrefix = {arXiv},
       eprint = {1402.5163},
 primaryClass = {astro-ph.IM},
       adsurl = {https://ui.adsabs.harvard.edu/abs/2014PASP..126..398H}
}

@ARTICLE{kup20,
       author = {{Kupfer}, Thomas and {Bauer}, Evan B. and {Marsh}, Thomas R. and
         {van Roestel}, Jan and {Bellm}, Eric C. and {Burdge}, Kevin B. and
         {Coughlin}, Michael W. and {Fuller}, Jim and {Hermes}, JJ and
         {Bildsten}, Lars and {Kulkarni}, Shrinivas R. and {Prince}, Thomas A. and
         {Szkody}, Paula and {Dhillon}, Vik S. and {Murawski}, Gabriel and
         {Burruss}, Rick and {Dekany}, Richard and {Delacroix}, Alex and
         {Drake}, Andrew J. and {Duev}, Dmitry A. and {Feeney}, Michael and
         {Graham}, Matthew J. and {Kaplan}, David L. and {Laher}, Russ R. and
         {Littlefair}, S.~P. and {Masci}, Frank J. and {Riddle}, Reed and
         {Rusholme}, Ben and {Serabyn}, Eugene and {Smith}, Roger M. and
         {Shupe}, David L. and {Soumagnac}, Maayane T.},
        title = "{The First Ultracompact Roche Lobe-Filling Hot Subdwarf Binary}",
      journal = {\apj},
         year = 2020,
        month = mar,
       volume = {891},
       number = {1},
          eid = {45},
        pages = {45},
          doi = {10.3847/1538-4357/ab72ff},
archivePrefix = {arXiv},
       eprint = {2002.01485},
 primaryClass = {astro-ph.SR},
       adsurl = {https://ui.adsabs.harvard.edu/abs/2020ApJ...891...45K}
}

@ARTICLE{pel21,
       author = {{Pelisoli}, Ingrid and {Neunteufel}, P. and {Geier}, S. and {Kupfer}, T. and {Heber}, U. and {Irrgang}, A. and {Schneider}, D. and {Bastian}, A. and {van Roestel}, J. and {Schaffenroth}, V. and {Barlow}, B.~N.},
        title = "{A hot subdwarf-white dwarf super-Chandrasekhar candidate supernova Ia progenitor}",
      journal = {Nature Astronomy},
         year = 2021,
        month = jul,
       volume = {5},
        pages = {1052-1061},
          doi = {10.1038/s41550-021-01413-0},
archivePrefix = {arXiv},
       eprint = {2107.09074},
 primaryClass = {astro-ph.SR},
       adsurl = {https://ui.adsabs.harvard.edu/abs/2021NatAs...5.1052P}
}

@ARTICLE{kup21,
       author = {{Kupfer}, Thomas and {Prince}, Thomas A. and {van Roestel}, Jan and {Bellm}, Eric C. and {Bildsten}, Lars and {Coughlin}, Michael W. and {Drake}, Andrew J. and {Graham}, Matthew J. and {Klein}, Courtney and {Kulkarni}, Shrinivas R. and {Masci}, Frank J. and {Walters}, Richard and {Andreoni}, Igor and {Biswas}, Rahul and {Bradshaw}, Corey and {Duev}, Dmitry A. and {Dekany}, Richard and {Guidry}, Joseph A. and {Hermes}, J.~J. and {Laher}, Russ R. and {Riddle}, Reed},
        title = "{Year 1 of the ZTF high-cadence Galactic plane survey: strategy, goals, and early results on new single-mode hot subdwarf B-star pulsatos}",
      journal = {\mnras},
         year = 2021,
        month = jul,
       volume = {505},
       number = {1},
        pages = {1254-1267},
          doi = {10.1093/mnras/stab1344},
archivePrefix = {arXiv},
       eprint = {2105.02758},
 primaryClass = {astro-ph.SR},
       adsurl = {https://ui.adsabs.harvard.edu/abs/2021MNRAS.505.1254K}
}

@ARTICLE{kup20a,
       author = {{Kupfer}, Thomas and {Bauer}, Evan B. and {Burdge}, Kevin B. and
         {Roestel}, Jan van and {Bellm}, Eric C. and {Fuller}, Jim and
         {Hermes}, JJ and {Marsh}, Thomas R. and {Bildsten}, Lars and
         {Kulkarni}, Shrinivas R. and {Phinney}, E.~S. and {Prince}, Thomas A. and
         {Szkody}, Paula and {Yao}, Yuhan and {Irrgang}, Andreas and
         {Heber}, Ulrich and {Schneider}, David and {Dhillon}, Vik S. and
         {Murawski}, Gabriel and {Drake}, Andrew J. and {Duev}, Dmitry A. and
         {Feeney}, Michael and {Graham}, Matthew J. and {Laher}, Russ R. and
         {Littlefair}, S.~P. and {Mahabal}, A.~A. and {Masci}, Frank J. and
         {Porter}, Michael and {Reiley}, Dan and {Rodriguez}, Hector and
         {Rusholme}, Ben and {Shupe}, David L. and {Soumagnac}, Maayane T.},
        title = "{A New Class of Roche Lobe-filling Hot Subdwarf Binaries}",
      journal = {\apjl},
         year = 2020,
        month = jul,
       volume = {898},
       number = {1},
          eid = {L25},
        pages = {L25},
          doi = {10.3847/2041-8213/aba3c2},
archivePrefix = {arXiv},
       eprint = {2007.05349},
 primaryClass = {astro-ph.SR},
       adsurl = {https://ui.adsabs.harvard.edu/abs/2020ApJ...898L..25K}
}

@ARTICLE{bau17,
   author = {{Bauer}, E.~B. and {Schwab}, J. and {Bildsten}, L.},
    title = "{Electron Captures on $^{14}$N as a Trigger for Helium Shell Detonations}",
  journal = {\apj},
archivePrefix = "arXiv",
   eprint = {1707.05394},
 primaryClass = "astro-ph.SR",
     year = 2017,
    month = aug,
   volume = 845,
      eid = {97},
    pages = {97},
      doi = {10.3847/1538-4357/aa7ffa},
   adsurl = {http://adsabs.harvard.edu/abs/2017ApJ...845...97B}
}

@ARTICLE{bro15,
   author = {{Brooks}, J. and {Bildsten}, L. and {Marchant}, P. and {Paxton}, B.
	},
    title = "{AM Canum Venaticorum Progenitors with Helium Star Donors and the Resultant Explosions}",
  journal = {\apj},
archivePrefix = "arXiv",
   eprint = {1505.05918},
 primaryClass = "astro-ph.SR",
     year = 2015,
    month = jul,
   volume = 807,
      eid = {74},
    pages = {74},
      doi = {10.1088/0004-637X/807/1/74},
   adsurl = {http://adsabs.harvard.edu/abs/2015ApJ...807...74B}
}

@INPROCEEDINGS{cor17,
       author = {{Cornish}, Neil and {Robson}, Travis},
        title = "{Galactic binary science with the new LISA design}",
    booktitle = {Journal of Physics Conference Series},
         year = 2017,
       series = {Journal of Physics Conference Series},
       volume = {840},
        month = may,
    publisher = {IOP},
          eid = {012024},
        pages = {012024},
          doi = {10.1088/1742-6596/840/1/012024},
archivePrefix = {arXiv},
       eprint = {1703.09858},
 primaryClass = {astro-ph.IM},
       adsurl = {https://ui.adsabs.harvard.edu/abs/2017JPhCS.840a2024C}
}

@ARTICLE{kup24,
       author = {{Kupfer}, Thomas and {Korol}, Valeriya and {Littenberg}, Tyson B. and {Shah}, Sweta and {Savalle}, Etienne and {Groot}, Paul J. and {Marsh}, Thomas R. and {Le Jeune}, Maude and {Nelemans}, Gijs and {Pala}, Anna F. and {Petiteau}, Antoine and {Ramsay}, Gavin and {Steeghs}, Danny and {Babak}, Stanislav},
        title = "{LISA Galactic Binaries with Astrometry from Gaia DR3}",
      journal = {\apj},
         year = 2024,
        month = mar,
       volume = {963},
       number = {2},
          eid = {100},
        pages = {100},
          doi = {10.3847/1538-4357/ad2068},
archivePrefix = {arXiv},
       eprint = {2302.12719},
 primaryClass = {astro-ph.SR},
       adsurl = {https://ui.adsabs.harvard.edu/abs/2024ApJ...963..100K}
}

@Article{cop10,
  Title                    = {{Physical properties of IP Pegasi: an eclipsing dwarf nova with an unusually cool white dwarf}},
  Author                   = {{Copperwheat}, C.~M. and {Marsh}, T.~R. and {Dhillon}, V.~S. and {Littlefair}, S.~P. and {Hickman}, R. and {G{\"a}nsicke}, B.~T. and {Southworth}, J.},
  Journal                  = {MNRAS},
  Year                     = {2010},

  Month                    = mar,
  Pages                    = {1824-1840},
  Volume                   = {402},
  Adsurl                   = {http://adsabs.harvard.edu/abs/2010MNRAS.402.1824C},
  Archiveprefix            = {arXiv},
  Doi                      = {10.1111/j.1365-2966.2009.16010.x},
  Eprint                   = {0911.1637},
  Primaryclass             = {astro-ph.SR}
}

@ARTICLE{fin10,
   author = {{Fink}, M. and {R{\"o}pke}, F.~K. and {Hillebrandt}, W. and 
	{Seitenzahl}, I.~R. and {Sim}, S.~A. and {Kromer}, M.},
    title = "{Double-detonation sub-Chandrasekhar supernovae: can minimum helium shell masses detonate the core?}",
  journal = {\aap},
archivePrefix = "arXiv",
   eprint = {1002.2173},
 primaryClass = "astro-ph.SR",
     year = 2010,
    month = may,
   volume = 514,
      eid = {A53},
    pages = {A53},
      doi = {10.1051/0004-6361/200913892},
   adsurl = {http://adsabs.harvard.edu/abs/2010A%26A...514A..53F}
}

@ARTICLE{sch19,
       author = {{Schaffenroth}, V. and {Barlow}, B.~N. and {Geier}, S. and {Vu{\v{c}}kovi{\'c}}, M. and {Kilkenny}, D. and {Wolz}, M. and {Kupfer}, T. and {Heber}, U. and {Drechsel}, H. and {Kimeswenger}, S. and {Marsh}, T. and {Wolf}, M. and {Pelisoli}, I. and {Freudenthal}, J. and {Dreizler}, S. and {Kreuzer}, S. and {Ziegerer}, E.},
        title = "{The EREBOS project: Investigating the effect of substellar and low-mass stellar companions on late stellar evolution. Survey, target selection, and atmospheric parameters}",
      journal = {\aap},
         year = 2019,
        month = oct,
       volume = {630},
          eid = {A80},
        pages = {A80},
          doi = {10.1051/0004-6361/201936019},
archivePrefix = {arXiv},
       eprint = {1907.09892},
 primaryClass = {astro-ph.SR},
       adsurl = {https://ui.adsabs.harvard.edu/abs/2019A&A...630A..80S}
}

@ARTICLE{gei13,
   author = {{Geier}, S. and {Marsh}, T.~R. and {Wang}, B. and {Dunlap}, B. and 
	{Barlow}, B.~N. and {Schaffenroth}, V. and {Chen}, X. and {Irrgang}, A. and 
	{Maxted}, P.~F.~L. and {Ziegerer}, E. and {Kupfer}, T. and {Miszalski}, B. and 
	{Heber}, U. and {Han}, Z. and {Shporer}, A. and {Telting}, J.~H. and 
	{G{\"a}nsicke}, B.~T. and {{\O}stensen}, R.~H. and {O'Toole}, S.~J. and 
	{Napiwotzki}, R.},
    title = "{A progenitor binary and an ejected mass donor remnant of faint type Ia supernovae}",
  journal = {\aap},
archivePrefix = "arXiv",
   eprint = {1304.4452},
 primaryClass = "astro-ph.SR",
     year = 2013,
    month = jun,
   volume = 554,
      eid = {A54},
    pages = {A54},
      doi = {10.1051/0004-6361/201321395},
   adsurl = {http://adsabs.harvard.edu/abs/2013A%26A...554A..54G}
}

@ARTICLE{heb86,
author = {{Heber}, U.},
title = "{The atmosphere of subluminous B stars. II - Analysis of 10 helium poor subdwarfs and the birthrate of sdB stars}",
journal = {\aap},
year = 1986,
month = jan,
volume = 155,
pages = {33-45},
adsurl = {http://adsabs.harvard.edu/abs/1986A%26A...155...33H}
}

@ARTICLE{heb16,
   author = {{Heber}, U.},
    title = "{Hot Subluminous Stars}",
  journal = {\pasp},
archivePrefix = "arXiv",
   eprint = {1604.07749},
 primaryClass = "astro-ph.SR",
     year = 2016,
    month = aug,
   volume = 128,
   number = 8,
    pages = {082001},
      doi = {10.1088/1538-3873/128/966/082001},
   adsurl = {http://adsabs.harvard.edu/abs/2016PASP..128h2001H}
}

@ARTICLE{heb09,
author = {{Heber}, U.},
title = "{Hot Subdwarf Stars}",
journal = {\araa},
year = 2009,
month = sep,
volume = 47,
pages = {211-251},
doi = {10.1146/annurev-astro-082708-101836},
adsurl = {http://adsabs.harvard.edu/abs/2009ARA%26A..47..211H}
}

@ARTICLE{ibe91,
   author = {{Iben}, Jr., I. and {Tutukov}, A.~V.},
    title = "{Helium star cataclysmics}",
  journal = {ApJ},
     year = 1991,
    month = apr,
   volume = 370,
    pages = {615-629},
      doi = {10.1086/169848},
   adsurl = {http://adsabs.harvard.edu/abs/1991ApJ...370..615I}
}

@ARTICLE{kup17a,
   author = {{Kupfer}, T. and {Ramsay}, G. and {van Roestel}, J. and {Brooks}, J. and 
	{MacFarlane}, S.~A. and {Toma}, R. and {Groot}, P.~J. and {Woudt}, P.~A. and 
	{Bildsten}, L. and {Marsh}, T.~R. and {Green}, M.~J. and {Breedt}, E. and 
	{Kilkenny}, D. and {Freudenthal}, J. and {Geier}, S. and {Heber}, U. and 
	{Bagnulo}, S. and {Blagorodnova}, N. and {Buckley}, D.~A.~H. and 
	{Dhillon}, V.~S. and {Kulkarni}, S.~R. and {Lunnan}, R. and 
	{Prince}, T.~A.},
    title = "{The OmegaWhite Survey for Short-period Variable Stars. V. Discovery of an Ultracompact Hot Subdwarf Binary with a Compact Companion in a 44-minute Orbit}",
  journal = {\apj},
archivePrefix = "arXiv",
   eprint = {1710.07287},
 primaryClass = "astro-ph.SR",
     year = 2017,
    month = dec,
   volume = 851,
      eid = {28},
    pages = {28},
      doi = {10.3847/1538-4357/aa9522},
   adsurl = {http://adsabs.harvard.edu/abs/2017ApJ...851...28K}
}

@ARTICLE{kup17,
   author = {{Kupfer}, T. and {van Roestel}, J. and {Brooks}, J. and {Geier}, S. and 
	{Marsh}, T.~R. and {Groot}, P.~J. and {Bloemen}, S. and {Prince}, T.~A. and 
	{Bellm}, E. and {Heber}, U. and {Bildsten}, L. and {Miller}, A.~A. and 
	{Dyer}, M.~J. and {Dhillon}, V.~S. and {Green}, M. and {Irawati}, P. and 
	{Laher}, R. and {Littlefair}, S.~P. and {Shupe}, D.~L. and {Steidel}, C.~C. and 
	{Rattansoon}, S. and {Pettini}, M.},
    title = "{PTF1 J082340.04+081936.5: A Hot Subdwarf B Star with a Low-mass White Dwarf Companion in an 87-minute Orbit}",
  journal = {\apj},
archivePrefix = "arXiv",
   eprint = {1612.02019},
 primaryClass = "astro-ph.SR",
     year = 2017,
    month = feb,
   volume = 835,
      eid = {131},
    pages = {131},
      doi = {10.3847/1538-4357/835/2/131},
   adsurl = {http://adsabs.harvard.edu/abs/2017ApJ...835..131K}
}

@ARTICLE{kup15a,
   author = {{Kupfer}, T. and {Geier}, S. and {Heber}, U. and {{\O}stensen}, R.~H. and 
	{Barlow}, B.~N. and {Maxted}, P.~F.~L. and {Heuser}, C. and 
	{Schaffenroth}, V. and {G{\"a}nsicke}, B.~T.},
    title = "{Hot subdwarf binaries from the MUCHFUSS project. Analysis of 12 new systems and a study of the short-period binary population}",
  journal = {A\&A},
archivePrefix = "arXiv",
   eprint = {1501.03692},
 primaryClass = "astro-ph.SR",
     year = 2015,
    month = apr,
   volume = 576,
      eid = {A44},
    pages = {A44},
      doi = {10.1051/0004-6361/201425213},
   adsurl = {http://adsabs.harvard.edu/abs/2015A\%26A...576A..44K}
}

@ARTICLE{bar22,
       author = {{Barlow}, Brad N. and {Corcoran}, Kyle A. and {Parker}, Isabelle M. and {Kupfer}, Thomas and {N{\'e}meth}, P{\'e}ter and {Hermes}, J.~J. and {Lopez}, Isaac D. and {Frondorf}, Will J. and {Vestal}, David and {Holden}, Jazzmyn},
        title = "{New Variable Hot Subdwarf Stars Identified from Anomalous Gaia Flux Errors, Observed by TESS, and Classified via Fourier Diagnostics}",
      journal = {\apj},
         year = 2022,
        month = mar,
       volume = {928},
       number = {1},
          eid = {20},
        pages = {20},
          doi = {10.3847/1538-4357/ac49f1},
archivePrefix = {arXiv},
       eprint = {2112.11463},
 primaryClass = {astro-ph.SR},
       adsurl = {https://ui.adsabs.harvard.edu/abs/2022ApJ...928...20B}
}

@ARTICLE{sav86,
   author = {{Savonije}, G.~J. and {de Kool}, M. and {van den Heuvel}, E.~P.~J.
	},
    title = "{The minimum orbital period for ultra-compact binaries with helium burning secondaries}",
  journal = {A\&A},
     year = 1986,
    month = jan,
   volume = 155,
    pages = {51-57},
   adsurl = {http://adsabs.harvard.edu/abs/1986A%26A...155...51S}
}

@ARTICLE{tut90,
   author = {{Tutukov}, A.~V. and {Yungelson}, L.~R.},
    title = "{Duplicity of Hot Helium Subdwarfs}",
  journal = {\sovast},
     year = 1990,
    month = feb,
   volume = 34,
    pages = {57},
   adsurl = {http://adsabs.harvard.edu/abs/1990SvA....34...57T}
}

@ARTICLE{tut89,
   author = {{Tutukov}, A.~V. and {Fedorova}, A.~V.},
    title = "{Formation and Evolution of Close Binary Stars Containing Helium Donors}",
  journal = {Soviet Astronomy},
     year = 1989,
    month = dec,
   volume = 33,
    pages = {606},
   adsurl = {http://adsabs.harvard.edu/abs/1989SvA....33..606T}
}

@ARTICLE{ven12,
   author = {{Vennes}, S. and {Kawka}, A. and {O'Toole}, S.~J. and {N{\'e}meth}, P. and 
	{Burton}, D.},
    title = "{The Shortest Period sdB Plus White Dwarf Binary CD-30 11223 (GALEX J1411-3053)}",
  journal = {\apjl},
archivePrefix = "arXiv",
   eprint = {1210.1512},
 primaryClass = "astro-ph.SR",
     year = 2012,
    month = nov,
   volume = 759,
      eid = {L25},
    pages = {L25},
      doi = {10.1088/2041-8205/759/1/L25},
   adsurl = {http://adsabs.harvard.edu/abs/2012ApJ...759L..25V}
}

@ARTICLE{wan18,
       author = {{Wang}, Bo},
        title = "{Mass-accreting white dwarfs and type Ia supernovae}",
      journal = {Research in Astronomy and Astrophysics},
         year = 2018,
        month = may,
       volume = {18},
       number = {5},
          eid = {049},
        pages = {049},
          doi = {10.1088/1674-4527/18/5/49},
archivePrefix = {arXiv},
       eprint = {1801.04031},
 primaryClass = {astro-ph.SR},
       adsurl = {https://ui.adsabs.harvard.edu/abs/2018RAA....18...49W}
}

@ARTICLE{wan12,
   author = {{Wang}, B. and {Han}, Z.},
    title = "{Progenitors of type Ia supernovae}",
  journal = {\nar},
archivePrefix = "arXiv",
   eprint = {1204.1155},
 primaryClass = "astro-ph.SR",
     year = 2012,
    month = jun,
   volume = 56,
    pages = {122-141},
      doi = {10.1016/j.newar.2012.04.001},
   adsurl = {http://adsabs.harvard.edu/abs/2012NewAR..56..122W}
}

@ARTICLE{yun08,
   author = {{Yungelson}, L.~R.},
    title = "{Evolution of low-mass helium stars in semidetached binaries}",
  journal = {Astronomy Letters},
archivePrefix = "arXiv",
   eprint = {0804.2780},
     year = 2008,
    month = sep,
   volume = 34,
    pages = {620-634},
      doi = {10.1134/S1063773708090053},
   adsurl = {http://adsabs.harvard.edu/abs/2008AstL...34..620Y}
}

@ARTICLE{bil07,
author = {{Bildsten}, L. and {Shen}, K.~J. and {Weinberg}, N.~N. and {Nelemans}, G. },
title = "{Faint Thermonuclear Supernovae from AM Canum Venaticorum Binaries}",
journal = {ApJL},
eprint = {arXiv:astro-ph/0703578},
year = 2007,
month = jun,
volume = 662,
pages = {L95-L98},
doi = {10.1086/519489},
adsurl = {http://adsabs.harvard.edu/abs/2007ApJ...662L..95B}
}

@ARTICLE{nel04,
author = {{Nelemans}, G. and {Yungelson}, L.~R. and {Portegies Zwart}, S.~F.},
title = "{Short-period AM CVn systems as optical, X-ray and gravitational-wave sources}",
journal = {MNRAS},
eprint = {arXiv:astro-ph/0312193},
year = 2004,
month = mar,
volume = 349,
pages = {181-192},
doi = {10.1111/j.1365-2966.2004.07479.x},
adsurl = {http://adsabs.harvard.edu/abs/2004MNRAS.349..181N}
}

@ARTICLE{han02,
   author = {{Han}, Z. and {Podsiadlowski}, P. and {Maxted}, P.~F.~L. and 
	{Marsh}, T.~R. and {Ivanova}, N.},
    title = "{The origin of subdwarf B stars - I. The formation channels}",
  journal = {MNRAS},
   eprint = {arXiv:astro-ph/0206130},
     year = 2002,
    month = oct,
   volume = 336,
    pages = {449-466},
      doi = {10.1046/j.1365-8711.2002.05752.x},
   adsurl = {http://adsabs.harvard.edu/abs/2002MNRAS.336..449H}
}

@ARTICLE{han03,
   author = {{Han}, Z. and {Podsiadlowski}, P. and {Maxted}, P.~F.~L. and 
	{Marsh}, T.~R.},
    title = "{The origin of subdwarf B stars - II}",
  journal = {MNRAS},
   eprint = {arXiv:astro-ph/0301380},
     year = 2003,
    month = may,
   volume = 341,
    pages = {669-691},
      doi = {10.1046/j.1365-8711.2003.06451.x},
   adsurl = {http://adsabs.harvard.edu/abs/2003MNRAS.341..669H}
}

@ARTICLE{nap04a,
   author = {{Napiwotzki}, R. and {Karl}, C.~A. and {Lisker}, T. and {Heber}, U. and 
	{Christlieb}, N. and {Reimers}, D. and {Nelemans}, G. and {Homeier}, D.
	},
    title = "{Close binary EHB stars from SPY}",
  journal = {Astrophysics and Space Science},
   eprint = {arXiv:astro-ph/0401201},
     year = 2004,
    month = jun,
   volume = 291,
    pages = {321-328},
      doi = {10.1023/B:ASTR.0000044362.07416.6c},
   adsurl = {http://adsabs.harvard.edu/abs/2004Ap%26SS.291..321N}
}

@ARTICLE{max01,
   author = {{Maxted}, P.~f.~L. and {Heber}, U. and {Marsh}, T.~R. and {North}, R.~C.
	},
    title = "{The binary fraction of extreme horizontal branch stars}",
  journal = {MNRAS},
   eprint = {arXiv:astro-ph/0103342},
     year = 2001,
    month = oct,
   volume = 326,
    pages = {1391-1402},
      doi = {10.1111/j.1365-8711.2001.04714.x},
   adsurl = {http://adsabs.harvard.edu/abs/2001MNRAS.326.1391M}
}

@ARTICLE{ma24,
       author = {{Ma}, Linhao and {Fuller}, Jim},
        title = "{Tidal Spin-up of Subdwarf B Stars}",
      journal = {\apj},
         year = 2024,
        month = nov,
       volume = {975},
       number = {1},
          eid = {1},
        pages = {1},
          doi = {10.3847/1538-4357/ad7788},
archivePrefix = {arXiv},
       eprint = {2408.16158},
 primaryClass = {astro-ph.SR},
       adsurl = {https://ui.adsabs.harvard.edu/abs/2024ApJ...975....1M}
}

@BOOK{sha83,
       author = {{Shapiro}, Stuart L. and {Teukolsky}, Saul A.},
        title = "{Black holes, white dwarfs and neutron stars. The physics of compact objects}",
         year = 1983,
          doi = {10.1002/9783527617661},
       adsurl = {https://ui.adsabs.harvard.edu/abs/1983bhwd.book.....S}
}

@BOOK{han04,
       author = {{Hansen}, Carl J. and {Kawaler}, Steven D. and {Trimble}, Virginia},
        title = "{Stellar interiors : physical principles, structure, and evolution}",
         year = 2004,
       adsurl = {https://ui.adsabs.harvard.edu/abs/2004sipp.book.....H}
}

@article{astropy,
doi = {10.3847/1538-4357/ac7c74},
url = {https://dx.doi.org/10.3847/1538-4357/ac7c74},
year = {2022},
month = {aug},
publisher = {The American Astronomical Society},
volume = {935},
number = {2},
pages = {167},
author = {Price-Whelan, Adrian M. and Lim, Pey Lian and Earl, Nicholas and Starkman, Nathaniel and Bradley, Larry and Shupe, David L. and Patil, Aarya A. and Corrales, Lia and Brasseur, C. E. and Nöthe, Maximilian and Donath, Axel and Tollerud, Erik and Morris, Brett M. and Ginsburg, Adam and Vaher, Eero and Weaver, Benjamin A. and Tocknell, James and Jamieson, William and van Kerkwijk, Marten H. and Robitaille, Thomas P. and Merry, Bruce and Bachetti, Matteo and Günther, H. Moritz and Paper Authors and Aldcroft, Thomas L. and Alvarado-Montes, Jaime A. and Archibald, Anne M. and Bódi, Attila and Bapat, Shreyas and Barentsen, Geert and Bazán, Juanjo and Biswas, Manish and Boquien, Médéric and Burke, D. J. and Cara, Daria and Cara, Mihai and Conroy, Kyle E and Conseil, Simon and Craig, Matthew W. and Cross, Robert M. and Cruz, Kelle L. and D’Eugenio, Francesco and Dencheva, Nadia and Devillepoix, Hadrien A. R. and Dietrich, Jörg P. and Eigenbrot, Arthur Davis and Erben, Thomas and Ferreira, Leonardo and Foreman-Mackey, Daniel and Fox, Ryan and Freij, Nabil and Garg, Suyog and Geda, Robel and Glattly, Lauren and Gondhalekar, Yash and Gordon, Karl D. and Grant, David and Greenfield, Perry and Groener, Austen M. and Guest, Steve and Gurovich, Sebastian and Handberg, Rasmus and Hart, Akeem and Hatfield-Dodds, Zac and Homeier, Derek and Hosseinzadeh, Griffin and Jenness, Tim and Jones, Craig K. and Joseph, Prajwel and Kalmbach, J. Bryce and Karamehmetoglu, Emir and Kałuszyński, Mikołaj and Kelley, Michael S. P. and Kern, Nicholas and Kerzendorf, Wolfgang E. and Koch, Eric W. and Kulumani, Shankar and Lee, Antony and Ly, Chun and Ma, Zhiyuan and MacBride, Conor and Maljaars, Jakob M. and Muna, Demitri and Murphy, N. A. and Norman, Henrik and O’Steen, Richard and Oman, Kyle A. and Pacifici, Camilla and Pascual, Sergio and Pascual-Granado, J. and Patil, Rohit R. and Perren, Gabriel I and Pickering, Timothy E. and Rastogi, Tanuj and Roulston, Benjamin R. and Ryan, Daniel F and Rykoff, Eli S. and Sabater, Jose and Sakurikar, Parikshit and Salgado, Jesús and Sanghi, Aniket and Saunders, Nicholas and Savchenko, Volodymyr and Schwardt, Ludwig and Seifert-Eckert, Michael and Shih, Albert Y. and Jain, Anany Shrey and Shukla, Gyanendra and Sick, Jonathan and Simpson, Chris and Singanamalla, Sudheesh and Singer, Leo P. and Singhal, Jaladh and Sinha, Manodeep and Sipőcz, Brigitta M. and Spitler, Lee R. and Stansby, David and Streicher, Ole and Šumak, Jani and Swinbank, John D. and Taranu, Dan S. and Tewary, Nikita and Tremblay, Grant R. and Val-Borro, Miguel de and Van Kooten, Samuel J. and Vasović, Zlatan and Verma, Shresth and de Miranda Cardoso, José Vinícius and Williams, Peter K. G. and Wilson, Tom J. and Winkel, Benjamin and Wood-Vasey, W. M. and Xue, Rui and Yoachim, Peter and Zhang, Chen and Zonca, Andrea and Astropy Project Contributors},
title = {The Astropy Project: Sustaining and Growing a Community-oriented Open-source Project and the Latest Major Release (v5.0) of the Core Package*},
journal = {The Astrophysical Journal}
}

@software{ccdproc,
  author       = {Matt Craig and
                  Steve Crawford and
                  Michael Seifert and
                  Thomas Robitaille and
                  Brigitta Sipőcz and
                  Josh Walawender and
                  Steve Crawford and
                  Zé Vinícius and
                  Joe Philip Ninan and
                  Michael Droettboom and
                  Timothy Ellsworth Bowers and
                  Jiyong Youn and
                  Yash Gondhalekar and
                  Erik Tollerud and
                  P. L. Lim and
                  E. M. Bray and
                  Yoonsoo P. Bach and
                  stottsco and
                  VSN Reddy Janga and
                  walkerna22 and
                  Hans Moritz Günther and
                  Evert Rol and
                  Jaime A. and
                  Larry Bradley and
                  Adrian Price-Whelan and
                  Christoph Deil and
                  Jenna Ryon and
                  Kelvin Lee and
                  Kyle Barbary and
                  Benjamin Weiner},
  title        = {astropy/ccdproc: 2.4.1},
  month        = may,
  year         = 2023,
  publisher    = {Zenodo},
  version      = {2.4.1},
  doi          = {10.5281/zenodo.7986923},
  url          = {https://doi.org/10.5281/zenodo.7986923},
}

@inproceedings{kry,
  title={Quanta image sensors for space applications},
  author={Krynski, Joanna and Bernard, Vivian and Lalucaa, Val{\'e}rian and Le Roche, Alexandre and Materne, Alex and Virmontois, C{\'e}dric and Goiffon, Vincent},
  booktitle={Advanced Photon Counting Techniques XIX},
  volume={13448},
  pages={103--116},
  year={2025},
  organization={SPIE}
}

@article{dhi,
  title={HiPERCAM: a quintuple-beam, high-speed optical imager on the 10.4-m Gran Telescopio Canarias},
  author={Dhillon, VS and Bezawada, N and Black, M and Dixon, SD and Gamble, T and Gao, X and Henry, DM and Kerry, P and Littlefair, SP and Lunney, DW and others},
  journal={Monthly Notices of the Royal Astronomical Society},
  volume={507},
  number={1},
  pages={350--366},
  year={2021},
  publisher={Oxford University Press}
}

@misc{glass,
	author = {Tyson B. Littenberg, Neil J. Cornish"},
	title = {GLASS},
	howpublished = {free software (Apache 2.0)},
	year = {2025}
}

@ARTICLE{liv90,
   author = {{Livne}, E.},
    title = "{Successive detonations in accreting white dwarfs as an alternative mechanism for type I supernovae}",
  journal = {ApJl},
     year = 1990,
    month = may,
   volume = 354,
    pages = {L53-L55},
      doi = {10.1086/185721},
   adsurl = {http://adsabs.harvard.edu/abs/1990ApJ...354L..53L}
}

\end{document}